\documentclass[a4paper,fleqn,usenatbib]{mnras}
\usepackage{ae,aecompl,graphicx,amsmath,amssymb}

\title[NGC\,4235 and NGC\,4594 with the GMRT]{A GMRT Study of Seyfert Galaxies NGC\,4235 \& NGC\,4594: Evidence of Episodic Activity ?}
\author[Kharb et al.]{
P. Kharb,$^{1}$\thanks{E-mail: kharb@iiap.res.in}
S. Srivastava,$^{2, 3}$ V. Singh,$^{2, 3}$
J. F. Gallimore,$^{4}$
C. H. Ishwara-Chandra,$^{2}$
\newauthor{Ananda Hota$^{5}$}
\\
$^{1}$Indian Institute of Astrophysics, II Block, Koramangala, Bangalore 560034, India\\
$^{2}$National Centre for Radio Astrophysics, Tata Institute of Fundamental Research, Post Bag 3, Ganeshkhind, Pune 411007, India\\
$^{3}$Astronomy and Astrophysics Division, Physical Research Laboratory, Ahmedabad 380009, India\\
$^{4}$Bucknell University, Department of Physics and Astronomy, Lewisburg, PA 17837, United States\\
$^{5}$UM-DAE Centre for Excellence in Basic Sciences, Vidyanagari, Mumbai 400098, India}

\date{Accepted XXX. Received YYY; in original form ZZZ}
\pubyear{2016}

\begin{document}
\label{firstpage}
\pagerange{\pageref{firstpage}--\pageref{lastpage}}
\maketitle

\begin{abstract}
Low frequency observations at 325 and 610 MHz have been carried out for two ``radio-loud'' Seyfert galaxies, NGC\,4235 and NGC\,4594 (Sombrero galaxy), using the Giant Meterwave Radio {Telescope} (GMRT). The 610~MHz total intensity and $325-610$~MHz spectral index images of NGC\,4235 tentatively suggest the presence of a ``relic'' radio lobe, most likely from a previous episode of AGN activity. This makes NGC\,4235 only the second known Seyfert galaxy after Mrk\,6 to show signatures of episodic activity. {\it Spitzer} and {\it Herschel} infrared spectral energy distribution (SED) modelling using the clumpyDREAM code predicts star formation rates (SFR) that are an order of magnitude lower than those required to power the radio lobes in these Seyferts ($\sim0.13-0.23$~M$_{\sun}$~yr$^{-1}$ compared to the required SFR of $\sim2.0-2.7$~M$_{\sun}$~yr$^{-1}$ in NGC\,4594 and NGC\,4235, respectively). This finding along with the detection of parsec and sub-kpc radio jets in both Seyfert galaxies, that are roughly along the same position angles as the radio lobes, strongly support the suggestion that Seyfert lobes are AGN-powered. SED modelling supports the ``true'' type 2 classification of NGC\,4594: this galaxy lacks significant dust obscuration as well as a prominent broad-line region. Between the two Seyfert galaxies, there is an inverse relation between their radio-loudness and Eddington ratio and a direct relation between their Eddington-scaled jet power and bolometric power. 
\end{abstract}


\section{Introduction} \label{intro}
Seyfert galaxies are one of the several manifestations of the active galactic nuclei (AGN) phenomena, where the primary source of the enormous nuclear energy is mass accretion onto a supermassive black hole. The presence or absence of broad bases to permitted emission lines in the optical spectra has lead to the classification of Seyfert type 1s and 2s, respectively \citep{Khachikian74}. Different orientations of the central engine with respect to our line of sight and the resulting obscuration due to a dusty torus, is generally accepted to be the primary explanation for the different Seyfert types, with type 1s being viewed nearly face-on \citep[a.k.a. the Unified Scheme;][]{Antonucci93}. Seyfert galaxies have traditionally been referred to as ``radio-quiet'' AGN, i.e., the ratio of their 5~GHz radio flux density to their $B$-band optical flux density ($R$), is typically $<10$ \citep{Kellermann89}. However, \citet{HoPeng01} have demonstrated that when the optical nuclear luminosity is extracted through high resolution observations with the {\it Hubble Space Telescope} (HST) and the galactic bulge contribution is properly accounted for, then the majority of Seyfert galaxies (belonging to the CfA and Palomar samples in Ho \& Peng's study) transition into the ``radio-loud" class. \citet{Kharb14} confirmed this trend in the Extended 12$\mu$m Seyfert sample. Additional criteria \citep[e.g.,][]{Panessa07} must therefore be adopted to differentiate between the radio emission from Seyferts and radio galaxies that typically exhibit $\sim$100~kpc scale radio jets. {Alternatively, more sources that straddle the divide between the ``radio-loud" and the ``radio-quiet" classes, must be looked at \citep[e.g.,][and references therein]{VSingh15,Mao15}: these sources may point towards a genuine continuum in properties between AGN sub-classes.}

Sensitive radio observations of Seyferts have indeed revealed the presence of kiloparsec-scale radio structures \citep[KSRs;][]{Baum93,Thean00}. \citet{Gallimore06} detected KSRs in $>44\%$ of Seyferts in the volume-limited CfA+12$\mu$m sub-sample, when observed with the most sensitive D-array configuration of the Very Large Array (VLA) at 5 GHz. \citet{Singh15} have found that $>43\%$ of Seyferts from a sample derived from sources detected in the 1.4~GHz VLA FIRST\footnote{Faint Images of the Radio Sky at Twenty-cm} and NVSS\footnote{NRAO VLA Sky Survey} surveys, possess KSRs. The resolution limits at the median redshifts of these surveys typically miss Seyferts with KSR extents $\lesssim5$ and $\lesssim10$ kpc for VLA D-array at 5~GHz and B-array at 1.4~GHz, respectively. Sensitive multi-frequency VLA observations have detected two pairs of radio lobes/KSRs, perpendicular to each other, in the ``radio-loud'' Seyfert galaxy, Mrk\,6 \citep{Kharb06}. (We will use KSRs and radio lobes interchangeably in the remaining text.) Very long baseline interferometry (VLBI) has revealed a high brightness temperature ($T_b\sim1\times10^8$\,K), inverted spectral index core and jet in Mrk\,6; this has similarly been observed in other Seyfert galaxies \citep[e.g.,][]{Falcke00,Ulvestad01,Kharb10,Kharb15}. These studies support the suggestion that radio {lobes} in Seyferts are AGN-driven and nonthermal (synchrotron) in origin. Additionally, Mrk\,6 shows signatures of episodic AGN activity.

In order to probe the full extent of the KSRs in two Seyfert galaxies {previously} known to possess KSRs, {\it viz.}, NGC\,4235 and NGC\,4594 \citep[see][]{Gallimore06}, and to ascertain their origin, we carried out low frequency observations with the Giant Meterwave Radio Telescope (GMRT) at 325 and 610~MHz. The results of this study are presented here. We also present results from an infrared (IR) spectral energy distribution (SED) modelling of NGC\,4235 and NGC\,4594. Our motivation was to detect any relic emission, if present, at the low radio frequencies of the GMRT, and to examine the relationship between radio {lobes} and galactic stellar and interstellar medium (ISM) components, which are probed by infrared observations. The ultimate goal is to disentangle the role of the AGN versus the stellar component in producing radio outflows in Seyfert galaxies.

For this paper, we adopt a cosmology where H$_0$ = 73~km~sec$^{-1}$~Mpc$^{-1}$, $\Omega_{m}$ = 0.27, $\Omega_{v}$ = 0.73. Spectral index $\alpha$ is defined such that flux density at a frequency $\nu$ is $S_\nu\propto\nu^\alpha$.

\section{Data and Analysis} \label{data}
\subsection{Sources of Study}
NGC\,4235 is a nearby (redshift, $z$ = 0.008039, scale = 0.181~kpc~arcsec$^{-1}$) Seyfert 1.2 galaxy residing in a nearly edge-on SA(s)a type spiral host {with no stellar bar or ring. Seyfert type 1.2 galaxies have ratios of total H$\beta$ to [OIII] $\lambda$5007 fluxes varying between 2.0 and 5.0 \citep{Winkler92,Veron06}. }
{NGC\,4594 is the famous Sombrero galaxy lying at the redshift of $z$ = 0.003416 (scale = 0.090~kpc~arcsec$^{-1}$). It is also an edge-on SA(s)a type spiral galaxy with a LINER\footnote{Low-ionization nuclear emission-line region} nucleus \citep{Ho97}. It has alternatively been classified as an unobscured ``true'' type 2 AGN: such objects have a weak or missing broad line region \citep[BLR;][]{Panessa02,Shi10}.}

\subsection{Radio Data}
Radio observations of NGC\,4235 and NGC\,4594 at 325 and 610 MHz were carried out in 2012 June (project code 22\_066) with the GMRT \citep{Swarup91}. 30 antennas at bandwidths of 32 MHz at both frequencies were used. A journal of these observations is presented in Table 1. The amplitude calibrators were observed for 20 min at the start and/or end of each run, and the phase calibrators were observed for 5 min every 40 min. NGC\,4235 and NGC\,4594 were observed for approximately four hours at each frequency.

\begin{table}
\centering
\small
\caption{Details of the GMRT Observations}
\begin{tabular}{@{}llllll@{}}
\hline\hline
Source                                & \multicolumn{2}{c}{NGC\,4594}     & \multicolumn{2}{c}{NGC\,4235}    \\\hline
Frequency                  &  325 MHz     &  610 MHz      &  325 MHz     &  610 MHz       \\
Date                   & $06/06/2012$     & $11/06/2012$      & $15/07/2012$     & $20/06/2012$ \\
Amp Cal           &  3C\,286         &  3C\,147, 3C\,286   & 3C\,147, 3C\,286   & 3C\,286   \\
$S_\nu$ (Jy)$^{\dagger}$    &  26.4          &  36.6, 21.3     & 55.8, 24.2     & 22.0 \\
Phase Cal          & 1311$-$222     &  1311$-$222     & 1123+055     &  1120+143     \\
$S_\nu$ (Jy)$^{\ddagger}$   & 25.08$\pm$0.16 &  1.8$\pm$0.03   & 4.63$\pm$0.10  & 3.65$\pm$0.04\\
\hline
\end{tabular}\\
{Column 1 lists the observing frequencies and dates. Amp Cal and Phase Cal refer to the amplitude and phase calibrators, respectively. $\dagger$ Flux density of amplitude calibrator set by the {\tt SETJY} task in {\tt AIPS}, using the (1999.2) VLA or Reynolds (1934-638) coefficients.
$\ddagger$ Flux density and error for the phase calibrators from {\tt AIPS} task {\tt GETJY}}
\label{tabobs}
\end{table}

Data were analysed using the NRAO Astronomical Image Processing System ({\tt AIPS}). Standard steps in data reduction were followed. The task {\tt TVFLG} was used to identify and remove the channels affected by radio frequency interference (RFI). Due to lower sensitivity and stability, the channels at the edges of the bands were also removed. Calibration using the primary and secondary calibrators was carried out after editing bad data. The edited and calibrated data on the target source were then averaged in frequency to an extent that kept the bandwidth smearing effect to negligible levels. The data were finally imaged using the wide-field imaging technique, to account for the non-coplanarity of the array. Several iterations of phase-only self-calibration and one iteration of amplitude and phase self-calibration were performed to improve the sensitivity of the images. The final 325 and 610 MHz images presented in Figures~\ref{fig1}-\ref{fig3} were created with a {\tt ROBUST} parameter of 0.25 and 2.0, respectively. The typical {\it rms} noise in the 610~MHz images was $\sim45~\mu$Jy~beam$^{-1}$ and in 325~MHz images was $\sim150~\mu$Jy~beam$^{-1}$. We find that the dynamic range of the images at 610~MHz was 1.5 to 3.0 times better than at 325~MHz.

The $325-610$ MHz spectral index and spectral index noise images were created using the {\tt AIPS} task {\tt COMB}, by first obtaining similar resolution images at both frequencies using appropriate weighting with the {\tt ROBUST} parameter, and subsequently convolving both images with identical circular beams in {\tt IMAGR} ($8\times8$~arcsec$^2$ beam for NGC\,4235, $10\times10$~arcsec$^2$ beam for NGC\,4594). Images were shifted to be exactly spatially coincident using the task {\tt OGEOM}. Regions with flux densities {below $1.5\sigma$} were blanked while using {\tt COMB}. 

\subsection{Infrared Data} \label{secir}
The nuclear structure of galaxies can be probed by infrared data. To this end, we gathered all archival IR data on the two Seyfert galaxies. We then decomposed the infrared spectra and spectral energy distributions using a customised version of the code clumpyDREAM (J. Gallimore, in preparation). clumpyDREAM makes the usual assumption that the infrared SEDs of galaxies can be decomposed into a sum of contributions from stars (photospheric emission), interstellar dust and PAHs (the ISM component), and re-radiation from the dusty torus surrounding an AGN \citep{Genzel00,Walcher11}. 

{The near-IR data for NGC\,4235 come from the 2MASS\footnote{Two Micron All Sky Survey \citep{Skrutskie06}} extended source catalog \citep{Jarrett00}, while the mid-IR} {\it Spitzer} IRS\footnote{InfraRed Spectrograph} data were obtained from \citet{Lebouteiller11}. These data were obtained in {\tt staring} rather than {\tt mapping} mode, and so the apertures vary: SL (short-low; $\sim5-14$ microns) slit = 3.6~arcsec; LL (long-low; $\sim14-38$ microns) slit = 10.5~arcsec. As a result, any resolved emission will contribute more to the LL module, resulting in a spectral jump at around 14 microns. clumpyDREAM fits for this jump assuming that the AGN components are unresolved but the stars and ISM are resolved. The 60 and 100 micron data were taken from the  IRAS\footnote{Infrared Astronomical Satellite} point source catalog, and the $250-500$ micron points were extracted from {\it Herschel} SPIRE\footnote{Spectral and Photometric Imaging Receiver} observations. 

The {\it Spitzer} IR data for NGC\,4594 come from \citet{Gallimore10}. The {\it Spitzer} IRS + IRAC\footnote{InfraRed Array Camera} spectra were extracted from 20~arcsec synthetic apertures. {The near-IR data from 2MASS are aperture-matched to the IRS extraction aperture.} The MIPS\footnote{Multiband Imaging Photometer for Spitzer} data were extracted from a 29~arcsec (parallel to slit) $\times$ 20~arcsec (slit width) aperture, assuming a point source geometry. The long-wavelength ($250-500$ micron) data come from archival {\it Herschel} SPIRE observations, extracted as background-subtracted point sources.

The clumpyDREAM code performs a Markov Chain Monte Carlo (MCMC) search over a parameter grid comprising of stellar, ISM and AGN torus components \citep[for {more} details see appendix in][]{Sales15}. The code also allows for contributions from very hot ($\sim1500$~K), presumably carbonaceous dust grains and AGN-heated dust in the narrow line region. 
The hot dust component is essentially a part of the dust distribution that the CLUMPY\footnote{http://www.pa.uky.edu/clumpy/} models of Heymann, Nikutta, and Elitzur, do not take into account \citep{Mor09}. {These} models assume the same sublimation radius for all grains, when in fact carbonaceous grains have a smaller sublimation radius than silicate grains. Several sources show that the hot dust component is reasonably well modelled by a single blackbody temperature \citep[e.g.,][]{Rodriguez06}. 

The torus clouds are distributed in angular altitude ($\theta$) above the torus equator as $N$($\theta$) = $N_0$ exp($-\theta^2 / \sigma^2$). From the point-of-view of the AGN or accretion disk, the torus covers a certain fraction of the sky (the "covering fraction" of the torus). Therefore, the integrated luminosity of the torus = (the AGN luminosity) $\times$ (the torus covering fraction). The torus luminosity and covering fraction are derived from the model fit, which in turn gives us the AGN bolometric luminosity. 

The stellar population is modelled as a single stellar population that undergoes continuous star-formation for 1 Gyr and then passively evolves afterward \citep{Silva98}. Star formation rate (SFR) is derived using the far-IR relation (equation 4) in \citet{Kennicutt98}. However, since this relation was calibrated for starbursts, there appears to be a $\approx30\%$ systematic uncertainty based on comparison to other calibrations. For more quiescent star-forming galaxies, like NGC\,4235 and NGC\,4594, this calibration may underestimate the SFR by as much as a factor of two. The ISM luminosity is calculated by numerically integrating the IR spectrum of the ISM component from $8-1000$ microns; it is not bolometric as it does not include UV/visible light that escapes dust extinction. 

{Upper and lower limits (from larger and smaller apertures) are necessary to provide the best possible constraints on model parameters. Looking at the far-IR region of the SED, fitting the \citet{Draine07} models strictly to the IRS spectra would result in acceptable models where the predicted far-IR SED would greatly exceed observational limits. There is a similar issue for the contribution of stellar photospheres at the near-IR region of the SED. IRS provides an anchor on starlight only at the shortest wavelengths ($\sim$5 microns), but stellar populations present a wide range of near-IR / 5 micron ratios, depending on the mix and age of the underlying populations. Upper limits are necessary to keep that ratio from running away, which then affects how the SED decomposes in the 5-10 micron range, above which dust clearly dominates. Note that we have matched aperture broadband measurements at near-mid IR wavebands (2MASS and IRAC) for NGC\,4594. In other words, the fit for NGC\,4594 is a fit to 2MASS + IRAC + IRS, but MIPS and Herschel data provide clamps on the far-IR contribution of the Draine \& Li models.}

\section{Results}\label{result}
In this Section, we present the results from the radio imaging, followed by those from the IR SED modelling. We first look at some global properties of NGC\,4235 and NGC\,4594. While Seyfert galaxies have historically been classified as ``radio-quiet'', we find that using the total 5~GHz flux density from \citet[][11.0 and 5.8~mJy for NGC\,4235 and NGC\,4594, respectively]{Gallimore06}, and ``nuclear'' optical B-band (4400\,\AA) flux density from the {HST} \citep[0.2~mJy and 72~$\mu$Jy for NGC\,4235 and NGC\,4594, respectively,][]{Papadakis08,Maoz05}, the radio-loudness parameter ($R$) turns out to be $\approx$50 for NGC\,4235 and $\approx$80 for NGC\,4594, making them both ``radio-loud'' in the conventional sense. We used the F606W/WFPC2/HST flux density for NGC\,4235 and the F330W/ACS/HST flux density for NGC\,4594. An optical spectral index of $-0.9$ \citep[e.g.,][]{Chiaberge99} was used to convert flux densities between different optical bands. 

\citet{Dong06} have estimated the black hole masses of NGC\,4235 and NGC\,4594 to be $\sim2.5\times10^7$M$_{\sun}$ and $\sim7.9\times10^8$\,M$_{\sun}$, respectively. They used the calibrated black hole - 2MASS K$_s$-band bulge luminosity relation to estimate the {same}. Using the [O {\sc iii}] line fluxes listed in \citet{Robitaille07} and \citet{Gonzalez09} for NGC\,4235 and NGC\,4594, respectively, and the formulation of \citet{Heckman04}, we {have} estimated the bolometric luminosities (L$_\mathrm{bol}$). These are tabulated in Table~\ref{tabbol}. In Section~\ref{discuss} ahead, we compare these to the more accurate bolometric luminosities obtained from the infrared SED modelling.

\subsection{Radio Lobes: Morphology, Energetics and Outflow Properties}\label{morph}
 The double-lobed structure previously observed by \citet{Gallimore06} is clearly seen in the 325 and 610 MHz images of NGC\,4235 (Figure~\ref{fig1}). A distinct ``hotspot'' is seen in the eastern lobe. This was not apparent in the previous VLA observations of \citet{Gallimore06}. The size of individual lobes from the galaxy center is $\approx12$~kpc; the total radio extent is $\approx24$~kpc. We also see a hint of diffuse emission at the $3\sigma$ intensity level, extending beyond the clearly-demarcated western lobe in the 610~MHz image of NGC\,4235. However, the higher noise levels in the 325~MHz image prevent us from observing the same (the $3\sigma$ intensity level of the 610~MHz image is similar to the $1\sigma$ level of the 325~MHz image). We discuss this diffuse emission ahead in Section~\ref{secrelic}. The $325-610$ MHz spectral index image of NGC\,4235 (Figure~\ref{fig2}) shows that {the} average spectral index is $-0.29\pm0.19$ in the western lobe and $-0.56\pm0.24$ in the eastern lobe.

The double lobes of NGC\,4594 are immersed in emission from the galactic disk, which gets more prominent at 325~MHz (Figure~\ref{fig3}). The higher frequency (5~GHz) VLA observations of \citet{Gallimore06} did not detect any galactic emission in this source.
{However, recent sensitive observations at 6~GHz with the Expanded VLA (EVLA), that achieve image {\it rms} noise levels of $\sim8~\mu$Jy~beam$^{-1}$, do detect the galactic emission in NGC\,4594 \citep{Wiegert15}.} 
The size of the south-eastern lobe in NGC\,4594 from the galaxy center is $\approx$3.5~kpc, the total radio extent of the lobes is $\approx$5.0 kpc. {These lobes are best seen in Figure~\ref{fig5}, where the brightest emission is shown and the disk emission is cut out of the frame.}
The average $325-610$~MHz spectral index in the southern lobe is inverted, $\alpha\sim0.29\pm0.16$ (Figure~\ref{fig4}). We note that the lobe flux density and spectral index values in NGC\,4594 are likely to be contaminated by its galactic emission. The galaxy disk spectral index is on average around $-1.20\pm0.25$. 

In addition, a ``spur'' of  extent $\sim3.0$ kpc is observed in the north-eastern direction at both the radio frequencies in NGC\,4594 (Figure~\ref{fig5}). The spectral index of the ``spur'' is difficult to extract in the GMRT image due to the presence of the diffuse galactic emission, but appears to be similar to the average spectral index of the lobes. The fact that this feature is not symmetrical about the core and is observed at both the frequencies, gives credence to its reality. The 5~GHz VLA image of \citet{Gallimore06} did not detect such a feature. Considering that the {\it rms} noise achieved at 5~GHz by \citet{Gallimore06} was $\sim50~\mu$Jy~beam$^{-1}$, and the intensity of the feature in our 610~MHz image is typically around 1~mJy~beam$^{-1}$, a $0.6-5.0$~GHz spectral index of $-0.9$ would be implied for this linear feature to fall below $3\sigma$ at 5~GHz, and therefore not be detected. This spectral index value is not unreasonable for optically thin synchrotron emission from a jet. Interestingly, the 1.5~GHz total and polarised intensity VLA images of NGC\,4594 by \citet{Bajaja88} do show the presence of this feature, although the authors do not comment on it. The ``spur'' that they refer to is instead the south-eastern lobe/jet that is observed by us and \citet{Gallimore06}. Assuming that the real ``spur'' intensity is fives times the {\it rms} noise in figure~2 of \citet{Bajaja88}, we derive a $0.6-1.5$~GHz spectral index of $-1.1$, which is again not unreasonable for a jet-like feature. We discuss this feature further in Section~\ref{discuss}. 

{Interestingly, the recent sensitive 1.5~GHz EVLA observations of \citet{Wiegert15} appear to show a broad northern ``plume'', which is much larger in extent than the ``spur'' and the northern lobe that we observe in the GMRT image, and is at an intermediate position angle (PA) in the sky. The poorer resolution of the EVLA D-array image ($\approx50$~arcsec) compared to the 610~MHz GMRT image ($\approx10$~arcsec) does not allow us to make a one-to-one comparison of the various features.}

We note that galactic emission was most likely not detected in NGC\,4235 at the low radio frequencies of the GMRT, due to its higher redshift compared to  NGC\,4594. Galactic emission in NGC\,4594 that is typically around 350~$\mu$Jy~beam$^{-1}$ at 610~MHz falls to 6~$\mu$Jy~beam$^{-1}$ at the redshift of NGC\,4235, which is much lower than the noise level at this frequency.

\subsubsection{``Relic'' Lobe Emission in NGC\,4235 (?)}\label{secrelic}
Both the 610~MHz total intensity image and the $325-610$ MHz spectral index image of NGC\,4235 hint at diffuse steep-spectrum radio emission just beyond the well-defined western lobe. This diffuse emission {\it envelopes} the western lobe: its shape resembles the shape of {the} western lobe and it has an extent that is roughly 1.5 times the extent of the western lobe. The average spectral index is $-1.82\pm0.17$ in this extended region. This surrounding steep-spectrum emission is clearly reminiscent of ``relic'' radio emission, as observed for instance, in the lobes of the radio galaxy 3C\,388 by \citet{Roettiger94}, where a lobe with {an average spectral index of $\sim-0.8$} is {\it surrounded} by steep-spectrum emission {of spectral index $\sim-1.5$} from a previous AGN activity episode. We note that in principle, a much larger extent of the relic lobe (than 1.5 times the present lobe) cannot be ruled out. However, this extended relic would then have an even steeper spectral index than $-1.8$, making it characteristically different from the relic emission we see, and therefore unlikely.

In order to check if the broken-ring-like region of average $325-610$ MHz spectral index $-1.8$ could be produced due to the noise peaks in the 325~MHz image, we carried out a simple exercise. We shifted the 325~MHz image successively by 25, 35 and 45 pixels in the X-direction, and created $325-610$~MHz spectral index images after aligning only pixel (1,1) in both the images, and not the source coordinates. This would superimpose the ``relic'' emission in the 610~MHz image with noise in the 325~MHz image. We found that where random noise peaks did produce regions of definite spectral index coinciding with the relic emission, the spectral index did vary between $-1.7$ and $-2.0$. However, in no case could we create the broken-ring-like region of average spectral index of $-1.8$ surrounding the western lobe, that is observed in Figure~\ref{fig2}. We therefore conclude that while the relic emission is close to the noise level at 325~MHz, its presence is tentatively suggested by its peculiar encompassing morphology. {However, by virtue of the 325~MHz relic emission being close to the noise level, the relic spectral index value of $-1.8$, must be regarded as an upper limit.}

{To the best of our knowledge, the detection of relic lobe emission makes NGC\,4235 only the second Seyfert galaxy after Mrk\,6 to exhibit multiple episodes of AGN activity. However, relic emission on the hundred-kiloparsec-scale has been observed in the radio galaxy NGC\,1167 \citep{Shulevski12} and a few other sources at low radio frequencies \citep[e.g.,][]{Hurley15,Brienza16}.}

We attempt to obtain the relative ages of the two lobes on the western side of NGC\,4235 by using the spectral-ageing formalism of \citet{Myers85}. Several assumptions have been made here. We adopt the average spectral index of the eastern lobe ($\alpha\sim-0.6$) for this calculation, and assume that the eastern and western lobes have similar ages. This is {a} reasonable {assumption} judging from their similar extents and morphology. The magnetic field is assumed to have the equipartition value of 5~$\mu$G (see Table~\ref{tabequip}). Furthermore, in order to use figure~2 from Myers \& Spangler, it is assumed that the $325-610$~MHz and the $1.4-4.9$~GHz lobe spectral indices are similar. All these assumptions make the {\it absolute} lobe age untrustworthy. We therefore only concern ourselves with the {\it relative} ages rather than the {\it absolute} ones. Using the {electron spectral index} $\gamma$=2.0 {(or $\alpha$=$-0.5$)} curve for the Kardashev-Pacholczyk (KP) model, we obtain an $X_0$=0.22 and $X_0$=0.91 for $\alpha$=$-0.6$ for the inner lobe and $\alpha$=$-1.8$ for the outer ``relic'' lobe, respectively. This translates to an age of $\approx$40~Myr for the inner lobe and $\approx$80~Myr for the outer ``relic" lobe. This suggests that the ``relic" radio lobe is two times older than the active lobe in NGC\,4235.

\subsubsection{Equipartition Estimates: B-fields and Electron lifetimes}\label{energetics}
Assuming ``equipartition'' of energy between relativistic particles and the magnetic field \citep{Burbidge59} in the radio lobes, we obtain the minimum pressure, and the particle energy (electrons and protons) at minimum pressure using the relations in \citet{OdeaOwen87}. The more sensitive 610~MHz images were used for estimating the equipartition values for individual radio lobes, assuming cylindrical symmetry. We did not make estimations for the short and not-clearly-demarcated north-western lobe of NGC\,4594. The total radio luminosity was estimated assuming that the radio spectrum extends from 10~MHz to 100~GHz with a spectral index of $\alpha$=$-0.6$. Furthermore, it was assumed that the ratio of the relativistic proton to relativistic electron energy was unity. Estimates for two extreme values of the volume filling factor, {\it viz.,} 1 and $10^{-3}$, were obtained and are presented in Table~\ref{tabequip}. The extremely small volume filling factor of $10^{-3}$ is justified if we assume that the lobe edges are the projections of precessing {jets} in Seyfert galaxies \citep[e.g., see Mrk\,6 in][]{Kharb06}. The equipartition magnetic field strength turns out to be $\sim4-30~\mu$G in NGC\,4235, and $\sim8-60~\mu$G in NGC\,4594. These parameters are similar to those obtained in other Seyfert galaxies \citep[e.g.,][]{Kharb06}. 

The lifetime of electrons in the radio component undergoing both synchrotron radiative and inverse Compton losses on cosmic microwave background (CMB) photons was estimated using the relation by \citet{vanderlaan69}, where the magnetic field was assumed to be the equipartition magnetic field and the break frequency was 1.6~GHz. The magnetic field equivalent to the radiation, which was assumed to be predominantly CMB photons, $B_{r}$, was estimated using the relation, $B_{r}\simeq4\times10^{-6}$(1+$z$)$^2$~G. The electron lifetimes are $\sim4-40$~Myr in the lobes of NGC\,4235 and $\sim2-20$~Myr in the (southern) lobe of NGC\,4594. We note again that galactic disk emission contaminates the flux density value for the lobe of NGC\,4594.

\subsubsection{Can the Radio Lobes be Powered by Stellar Winds ?}
If we assume that the radio lobe emission has a stellar-wind origin, we can derive the star formation rates  required to produce them. We estimated the SFR required to power the lobes using the total lobe luminosity at 610~MHz, and the relations in \citet{Condon92,Condon02} for a Salpeter initial mass function over the mass range $0.1<M/M_{\sun}<100$. These turn out to be $\sim2.7$~M$_{\sun}$~yr$^{-1}$ for NGC\,4235 and $\sim2.0$~M$_{\sun}$~yr$^{-1}$ for NGC\,4594, for stars more massive than 0.1~M$_{\sun}$. A radio spectral index of $-0.6$ and $+0.3$ was assumed for the lobes of NGC\,4235 and NGC\,4594, respectively, for the SFR estimation. 

While these SFR estimates are modest and comparable to those derived in the Milky Way galaxy \citep[e.g.,][]{Robitaille10}, they are still an order of magnitude larger than those derived using the {\it Spitzer} and {\it Herschel} IR data for NGC\,4235 and NGC\,4594 (Section~\ref{secsed}). This supports the suggestion that the radio lobe emission in these two Seyfert galaxies is powered by the AGN, rather than by stellar winds. The infrared SFR estimates are from a region only slightly larger than one GMRT beam, and do not include all of the emission from the host galaxy.

{Another indirect piece of evidence that favours the suggestion that the radio lobes are AGN-powered, is the ``oblique'' orientation of the lobes to the host galaxy major axes. \citet{Gallimore06} have shown that while the lobes can lie at all angles w.r.t. the host galaxy major axes, there {appears to be} a preference for position angle differences to lie between $40\degr-60\degr$. The PA difference is $\sim40\degr$ in NGC\,4235, and $\sim50\degr$ in NGC\,4594. Stellar winds, on the other hand, are typically aligned with the galaxy minor axes, where a pressure gradient in the ISM exists \citep{Colbert96}. This indicates that the radio lobes are ``directed'' flows, which can more easily be produced by AGN jets rather than nuclear starbursts.}

\subsubsection{Radio Lobes and Parsec / sub-Kpc Scale Jets }
If the Seyfert lobes are AGN-driven, it is essential to look for AGN outflows on parsec-scales in these sources to see if, and how, they align with the kpc-scale radio structures. NGC\,4235 has been imaged with VLBI at 8.4~GHz by \citet{Anderson04}. No clear jet-like extension is observed on milliarcsecond-scales. However, the lowest contour levels do indicate a slight extension at a PA of $\approx45\degr$, towards the north-west. The sub-arcsecond-scale VLA A-array 8.4~GHz image of \citet{Kukula95} clearly reveals a one-sided jet towards the north-west in NGC\,4235 at the same PA. The jet in NGC\,4235 therefore seems to point towards the northern edge of the north-western lobe. Similar parsec-scale-jet and kpc-scale-lobe-edge connections have been observed in other Seyfert galaxies like NGC\,6764 and Mrk\,6 \citep{Kharb06,Kharb14}. 

\citet{Hada13} and \citet{Mezcua14} have observed the parsec-scale jet in NGC\,4594 at multiple radio frequencies. A clear S-shaped or wiggly jet with counterjet is observed in the 5 and 8.4 GHz VLBI images of \citet{Hada13}. The position angle of the parsec-scale jet-counterjet structure is similar to that of the kpc-scale radio lobes in the north-west $-$ south-east direction (Figure~\ref{fig3}). \citet{Hada13} deduce that the northern jet is the approaching one. The parsec-scale jet inclination ($i_j$) lies in the range $1\degr<i_j<25\degr$ and the jet speed ($v_j$) is $0.02c<v_j<0.2c$. The parsec-scale jet is therefore sub-relativistic and inclined at a small angle to our line of sight. 

The close alignment between the parsec- and sub-kpc-scale radio jets and lobes could in principle suggest that AGN jets are feeding the radio lobes. The non-alignment of the jet and the centres of the lobes (as opposed to the lobe edges) in NGC\,4235, could suggest that the jet is curved or S-shaped, as is also observed in Mrk\,6 \citep{Kharb14}. The small angle between the S-shaped jet in NGC\,4594 and its torus inclination derived in Section~\ref{secsed} {(torus inclination $\sim17\degr$)}, could further imply that the accretion disk, which is presumably the base of the jet, is warped. Warped accretion disks have been suggested to produce precessing jets \citep[e.g.,][]{Pringle97}.

\subsubsection{Lobe Speeds and Kinetic Power}
While there is no direct way to measure lobe advance speeds, they can be examined indirectly. Based on the detection statistics of KSRs ($\ge44\%$), which constrain the AGN duty-cycle in Seyferts, \citet{Gallimore06} have derived relations for the lifetimes and speeds of radio lobes. Using these relations, we derive for $n$ = 2 (two episodes of AGN activity) and a lobe length of 12 kpc in NGC\,4235, the duration of a single outflow, $\tau_{\mathrm{flow}}$ to be $\sim$20 Myr, and the lobe advance speed $V_{\mathrm{flow}}$, to be $\sim$600~km~s$^{-1}$. Interestingly, the lobe advance speed is of the same order as for shock-excited gas inferred from spectroscopic studies of the NLR in Seyfert galaxies \citep[e.g.,][]{Bicknell98}. We note that the outflow duration is about half the age we derived in Section~\ref{secrelic}. However, (1) assuming only a slightly different equipartition magnetic field of 7~$\mu$G (instead of 5~$\mu$G) for the lobe age calculation in Section~\ref{secrelic}, gives consistent lobe ages ($\approx20$~Myr) from both calculations, (2) the {\it absolute} lobe age derived in Section~\ref{secrelic} is unreliable due to the several assumptions made, and (3) the simple relation of \citet{Gallimore06} assumes that all episodes have identical outflow speeds, which may or may not be true.

{Lower limits on outflow speeds can also be derived by dividing the radial extent of the lobes by the electron lifetimes \citep{Baum93}. The electron lifetimes from synchrotron radiative losses (assuming ``equipartition'') and inverse Compton losses due to the CMB background, are presented in Table~\ref{tabequip}. Therefore, assuming a volume filling factor of unity for the equipartition estimate of the magnetic field, the lower limits on outflow speeds are $\sim200-250$~km~s$^{-1}$ for the east and west lobes of NGC\,4235, and $\sim150$~km~s$^{-1}$ for the south lobe of NGC\,4594. }

We estimate the time-averaged kinetic power ($Q$) of the outflows in the two Seyfert galaxies by using the total 610~MHz flux density of the lobes (without the cores) and the relation from \citet{Punsly11}, which has been proposed for powerful radio galaxies. \citet{Punsly11} have slightly modified the relation of \citet{Willott99} and used an uncertainty parameter $f$ of 15. 
The $f$ parameter includes uncertainty due to minimum energy departures, geometric effects, filling factors, protonic contribution and the low frequency cutoff.
We derive the 151~MHz flux density from the 610~MHz flux density assuming a lobe spectral index of $\alpha$=$-0.6$ for NGC\,4235 and $\alpha$=$+0.3$ for NGC\,4594. We obtain a jet kinetic power of $Q\approx3.2\times10^{40}$~ergs~s$^{-1}$ for NGC\,4235 and $Q\approx2.1\times10^{39}$~ergs~s$^{-1}$ for NGC\,4594. 

Using the alternate relation $Q$=$\frac{\mathrm{E_{min}}}{\eta~\times~\mathrm{Lobe~Age}}$ \citep{Rawlings91}, we obtain the same $Q$ estimate (= 3.2\,$\times\,10^{40}$~ergs~s$^{-1}$) for NGC\,4235. For this we have used the {minimum energy,} E$_\mathrm{min}$ = 2\,$\times\,10^{55}$ ergs (Table~\ref{tabequip}; a representative value was chosen for an intermediate filling factor between 1 and $10^{-3}$), lobe age = 40 Myr (see Section~\ref{secrelic}) and $\eta$ = 0.5 \citep[parameter that allows for work done on external medium;][]{Rawlings91}. Kinetic powers {of the order of} $\sim10^{39}-10^{40}$~ergs~s$^{-1}$ have been found in other low-luminosity AGN with direct estimates of mass outflow rates \citep[e.g., NGC\,1433,][]{Combes13}. This gives us confidence that the relations for time-averaged kinetic power derived for powerful radio galaxies, also seem to work in Seyfert galaxies with radio lobes.

\subsection{Nuclear Properties from the IR SED Modelling}\label{secsed}
The infrared SED modelling with the customised clumpyDREAM code decomposes the stellar and AGN contribution to the nuclear emission in NGC\,4235 and NGC\,4594. Modelling reveals that the best-fit model for NGC\,4594 includes (i) a clumpy dusty torus, (ii) diffuse ISM, and (iii) a stellar population. The best-fit model for NGC\,4235, on the other hand, includes these three components {plus} an additional, presumably carbonaceous, hot dust component with temperature $\sim1500$~K, that is described in Section~\ref{secir}. The hot dust component is found only in quasi-stellar objects (QSOs) and type 1 AGN. The {SED} best-fit models are presented in Figures~\ref{fig5}, \ref{fig6} and Tables~\ref{tabsed1}, \ref{tabsed2}. Estimators of the probability densities of the fitted parameters for both Seyfert galaxies are shown in Figures~\ref{figa}-\ref{figd}. The model parameters are described in the caption to Table~\ref{tabsed1}. The jump in slit width between the short-low SL and the long-low LL modules mentioned in Section~\ref{secir}, can be seen in the model spectrum of NGC\,4235 (Figure~\ref{fig5}). For NGC\,4235, it looked like the stars + ISM needed to drop by a factor of $\sim$23\% in the SL mode relative to the LL mode {(the `SL apcorr' parameter in Table~\ref{tabsed1})}.

Some relevant parameters from the model-fitting for NGC\,4235 are:  SFR = 0.23~M$_{\sun}$~yr$^{-1}$, dusty torus size = 0.5~parsec, torus mass = 0.3~M$_{\sun}$, torus inclination = 4$\degr$ (consistent with its type 1 classification). For NGC\,4594 we obtain: SFR = 0.13~M~$_{\sun}$~yr$^{-1}$, torus size = 0.7~parsec, torus mass = 0.008~M$_{\sun}$, torus inclination = 17$\degr$. The torus mass seems to be extremely low in NGC\,4594. However, since the AGN bolometric luminosity is also low ($\sim1.4\times10^{42}$~ergs~s$^{-1}$; Table~\ref{tabsed2}), the dust sublimation radius is smaller as well \citep{Mor09}. This implies that a torus that is not geometrically large and has low mass, can still obscure AGN emission. The torus in NGC\,4594 is however, nearly face-on. Dust obscuration must therefore not be significant. This would then be consistent with its classification as a ``true'' type 2 AGN. Finally, the SED modelling also determines the ``escape probability" for the accretion disk and BLR photons to escape the torus, to be essentially 100\% in both NGC\,4235 and NGC\,4594, making them both type 1 ({in the sense of being} unobscured) in the Unified scheme. 

The bolometric AGN luminosity derived from the IR SED modelling, which relies on the torus luminosity and its covering fraction, is $3.5\times10^{42}$~ergs~s$^{-1}$ for NGC\,4235 and $1.4\times10^{42}$~ergs~s$^{-1}$ for NGC\,4594. As discussed ahead in Section~\ref{discuss}, these estimates are more accurate than the ones derived from the  [O {\sc iii}] line fluxes in Section~\ref{result}, and are used for estimating relevant parameters like the Eddington ratio ($\lambda_\mathrm{Edd}\equiv$L$_\mathrm{bol}$/L$_\mathrm{Edd}$, where the Eddington luminosity is L$_\mathrm{Edd}\equiv1.25\times10^{38}$~M$_\mathrm{BH}$/M$_{\sun}$), and kinetic to bolometric power (or jet to accretion efficiency; see Table~\ref{tabbol}).

\subsubsection{Global Star Formation Rates from IRAS and Radio Data}
We can derive ``global'' star formation rates using infrared fluxes from IRAS. We estimated the global IR luminosity, L($8-1000$ microns), using \citet{Sanders96}, and subtracted off the AGN L($8-1000$ microns) component based on the SED model fit. The AGN-subtracted, IRAS estimates of the global IR luminosity in ergs~s$^{-1}$ are: log~L$_\mathrm{IR}$ = 42.30 to 42.99 for NGC\,4235 and 43.385$\pm$0.012 for NGC\,4594. Since NGC\,4235 does not have 12 and 25 micron detections but only upper limits, the values for luminosity and star formation rates are ranges rather than measured values with uncertainty. Using the relations in \citet{Kennicutt98}, we obtain global SFR values of $0.089-0.438~M_{\sun}$~yr$^{-1}$ for NGC\,4235 and $1.09\pm0.30~M_{\sun}$~yr$^{-1}$ for NGC\,4594. {Keeping in mind the factor-of-two systematic uncertainty mentioned earlier in Section~\ref{secir}, the upper SFR limit for NGC\,4235 is 0.88~M$_{\sun}$~yr$^{-1}$.} Overall, these global values are greater than the nuclear values {estimated in Section~\ref{secsed}}. This is not surprising since these galaxies are disky and the global values include all of the disk.

Using the 610~MHz flux density from the galactic disk in NGC\,4594 (derived by subtracting the core+lobe flux density from the total observed), we derive an SFR of $\sim0.82$~M$_{\sun}$~yr$^{-1}$ for stars more massive than 0.1~M$_{\sun}$. (A spectral index of $-1.2$ was assumed for the disk emission; see Section~\ref{energetics}.) This radio estimate is consistent with the global SFR from the IRAS data, given the uncertainties. A similar analysis cannot be carried out for NGC\,4235 because galactic disk emission is not detected at either of the GMRT frequencies. We conclude that the global SFR estimates are consistent, whether using radio or infrared fluxes. {This strongly suggests that the difference in the ``nuclear'' SFR estimates between the infrared and the radio, is significant.}

\section{Discussion} \label{discuss}
\begin{figure}
\centering
\includegraphics[width=12cm,trim=50 220 50 220]{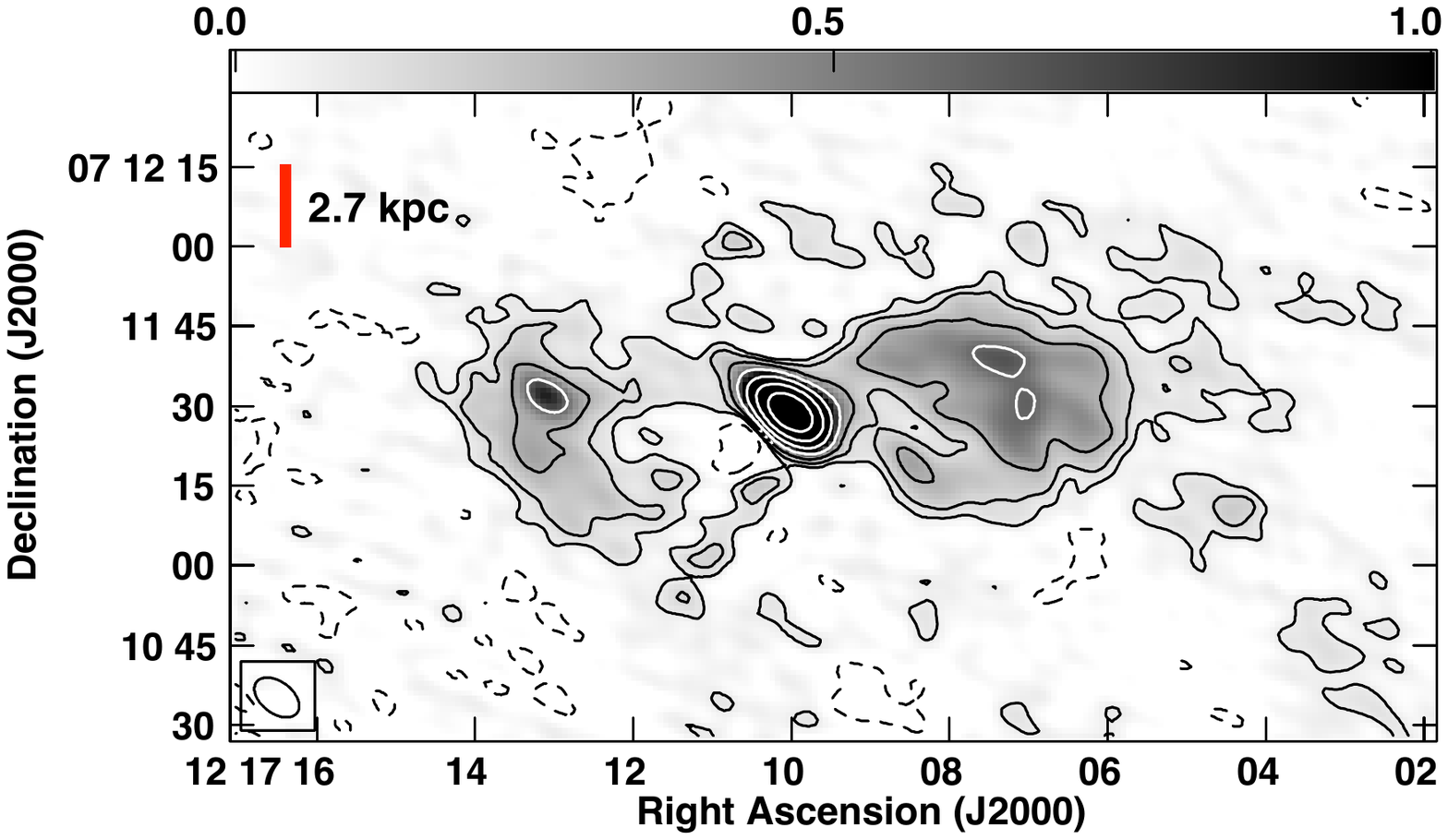}
\includegraphics[width=12cm,trim=50 220 50 220]{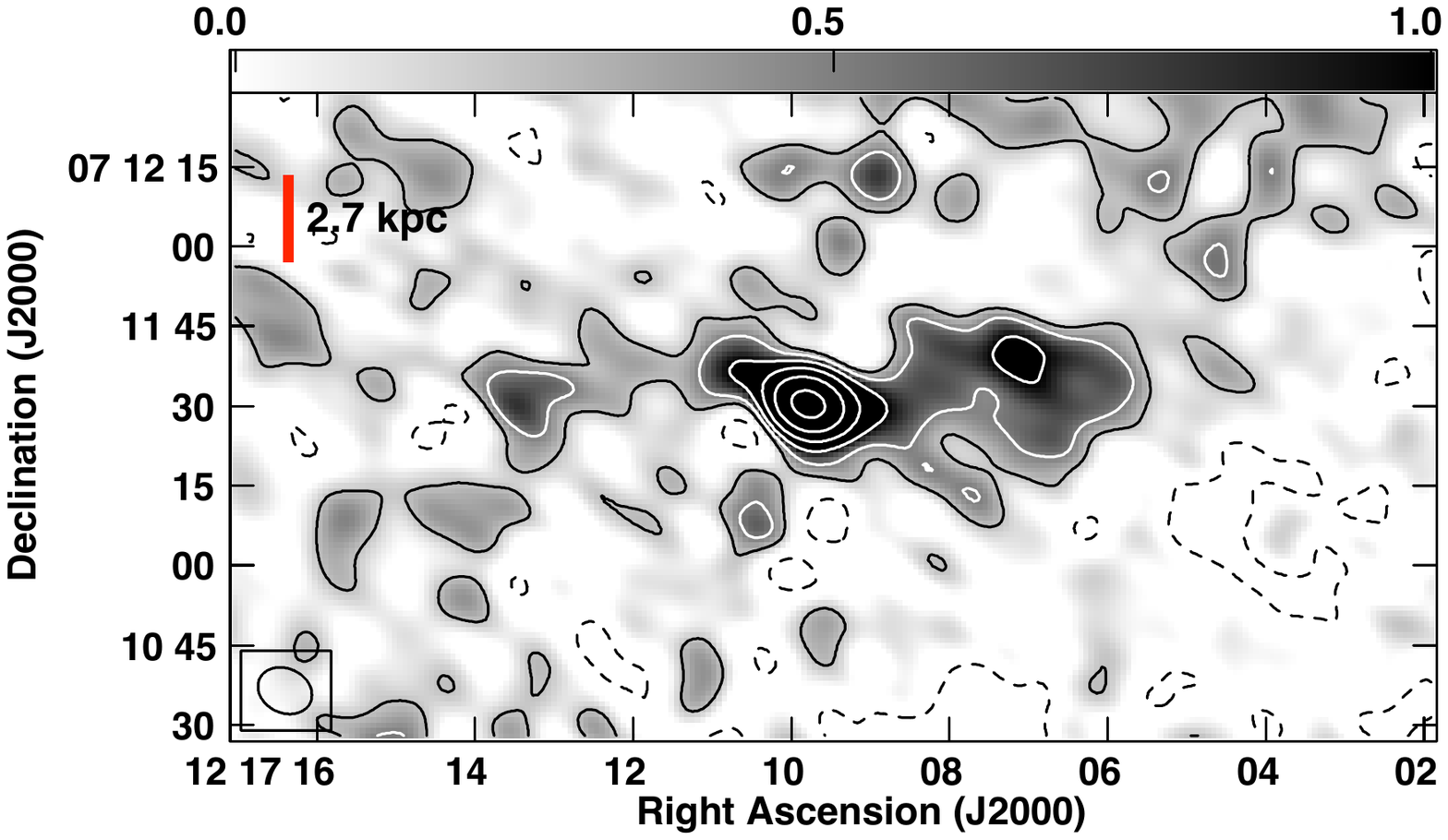}
\caption{610 MHz (top) and 325 MHz (bottom) contour images of NGC\,4235. The contour levels are in percentage of the peak intensity and increase {by factors of two}. The peak intensity and lowest contour levels are $2.5\times10^{-2}$~Jy~beam$^{-1}$ and $\pm$0.3\% for the top and $3.4\times10^{-1}$~Jy~beam$^{-1}$ and $\pm$0.074\% for the bottom panel, respectively. The beam-size is $9.6\times5.9$~arcsec$^2$ at a PA of $53\degr$ at 610 MHz, and $10.6\times8.3$~arcsec$^2$ at a PA of $65\degr$ at 325~MHz. The grey scale range is 0$-$1.0 mJy~beam$^{-1}$.}
\label{fig1}
\end{figure}

\begin{figure}
\centering
\includegraphics[width=12cm,trim=80 240 80 240]{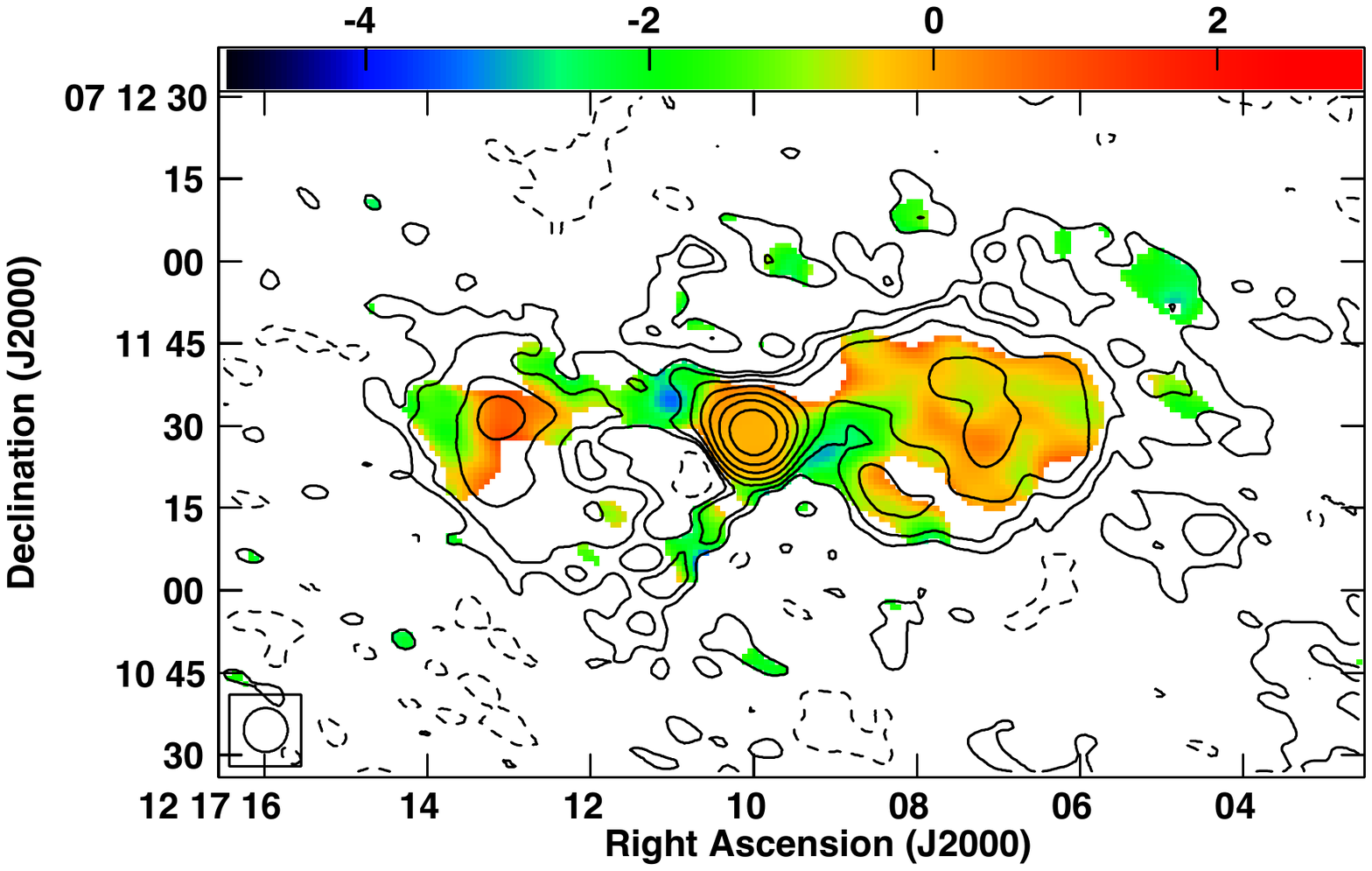}
\caption{$325-610$~MHz spectral index images in colour overlaid with 610 MHz radio contours for NGC\,4235. The average spectral index is $\sim-0.6$ in the lobes {and $\le-1.8$ in the ``relic'' lobe}. The contour levels are in percentage of the peak intensity and increase {by factors of two}. The peak intensity and lowest contour levels are $2.3\times10^{-2}$~Jy~beam$^{-1}$ and $\pm$0.3\%, respectively. The contour image is convolved with a circular beam of size 8~arcsec.}
\label{fig2}
\end{figure}

\begin{figure}
\centering
\includegraphics[width=12cm,trim=60 250 60 240]{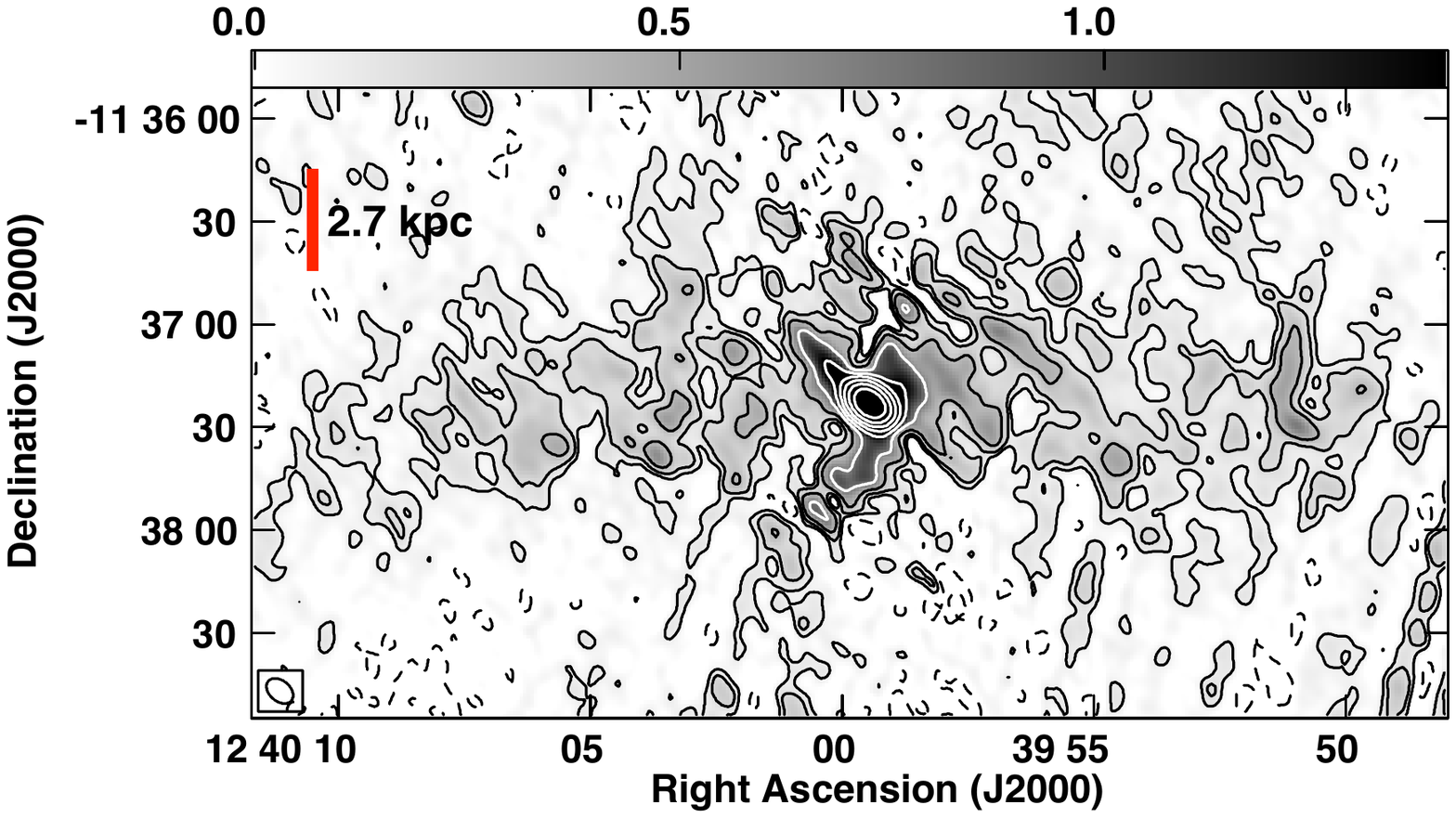}
\includegraphics[width=12cm,trim=60 240 60 250]{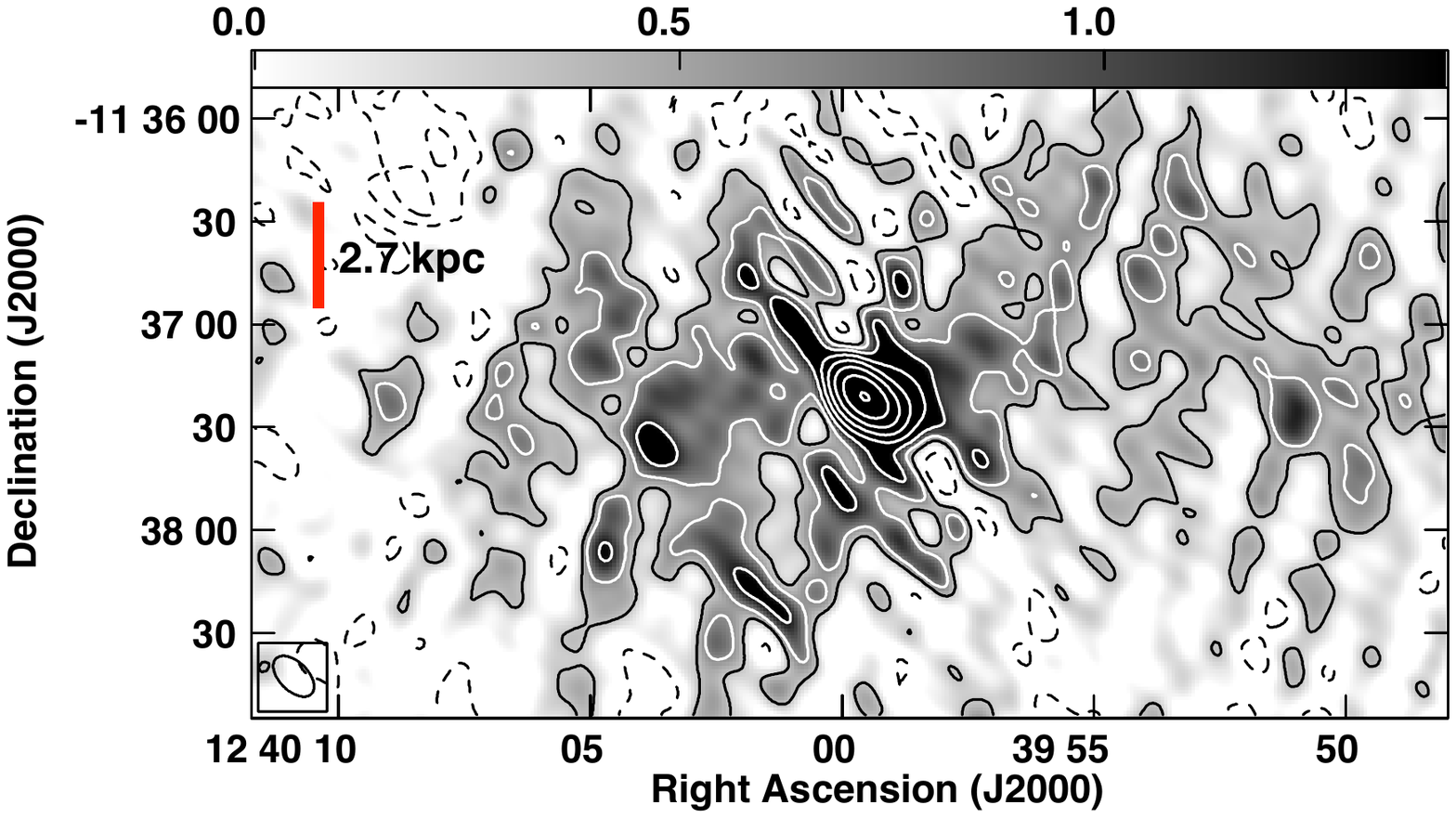}
\caption{610 MHz (top) and 325 MHz (bottom) contour images of NGC\,4594. The contour levels are in percentage of the peak intensity and increase {by factors of two}. The peak intensity and lowest contour levels are $6.1\times10^{-2}$~Jy~beam$^{-1}$ and $\pm$0.15\% for the top and $2.1\times10^{-1}$~Jy~beam$^{-1}$ and $\pm$0.15\% for the bottom panel, respectively. The beam-size is $8.8\times6.0$~arcsec$^2$ at a PA of $50\degr$ at 610 MHz, and $14.6\times8.4$~arcsec$^2$ at a PA of $45\degr$ at 325~MHz. The grey scale range is 0$-$1.4 mJy~beam$^{-1}$.} 
\label{fig3}
\end{figure}

\begin{figure}
\centering
\includegraphics[width=12cm,trim=80 240 80 240]{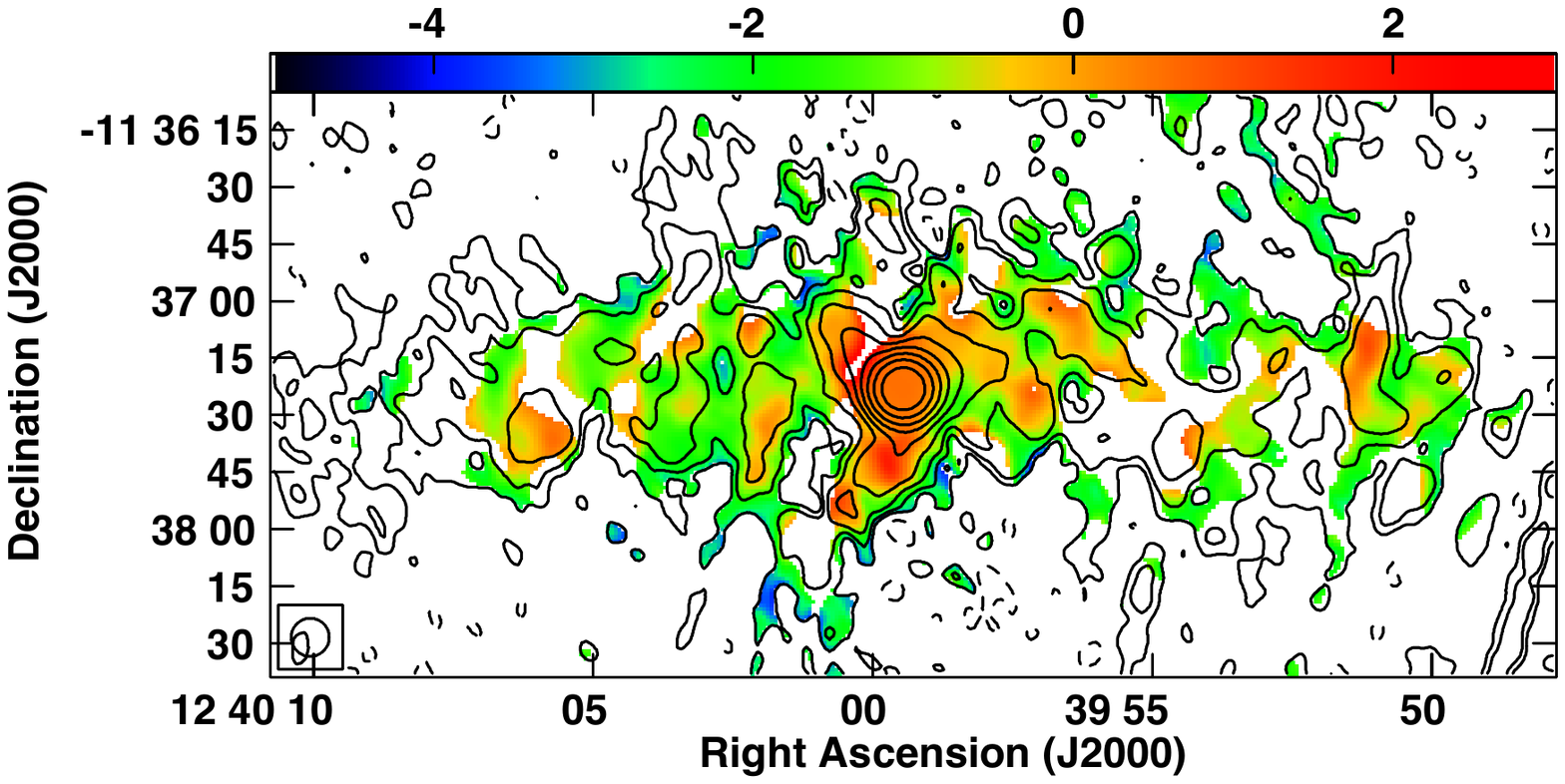}
\caption{$325-610$~MHz spectral index images in colour overlaid with 610 MHz radio contours for NGC\,4594. The average spectral index is +0.3 for the southern lobe of NGC\,4594. This value is however, likely to contaminated by galactic emission. The galaxy disk spectral index is $\sim-1.2$. The contour levels are in percentage of the peak intensity and increase {by factors of two}. The peak intensity and lowest contour levels are $6.7\times10^{-2}$~Jy~beam$^{-1}$ and $\pm$0.15\%, respectively. The contour image is convolved with a circular beams of size 10~arcsec.}
\label{fig4}
\end{figure}

\begin{figure}
\centering
\includegraphics[width=6cm,trim=100 100 100 100]{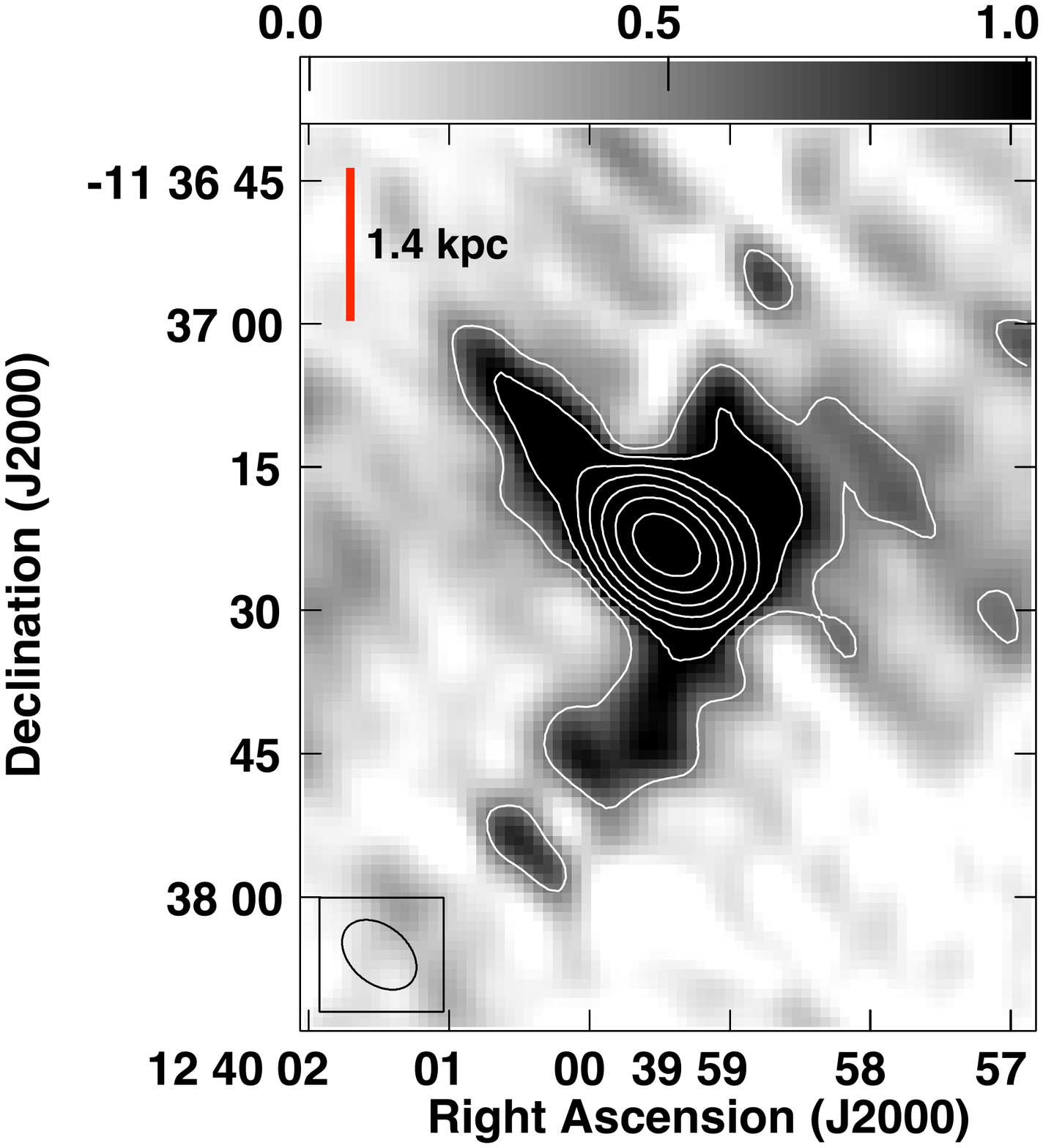}
\caption{The 610 MHz contour image of NGC\,4594, showing the ``spur" in the north-east direction. The contour levels are in percentage of the peak intensity and increase {by factors of two}. The peak intensity and lowest contour levels are $6.1\times10^{-2}$~Jy~beam$^{-1}$ and $\pm$0.85\%, respectively. {The beam-size is $8.8\times6.0$~arcsec$^2$ at a PA of $50\degr$.}}
\label{fig5}
\end{figure}

\begin{figure}
\centering
\includegraphics[width=18cm]{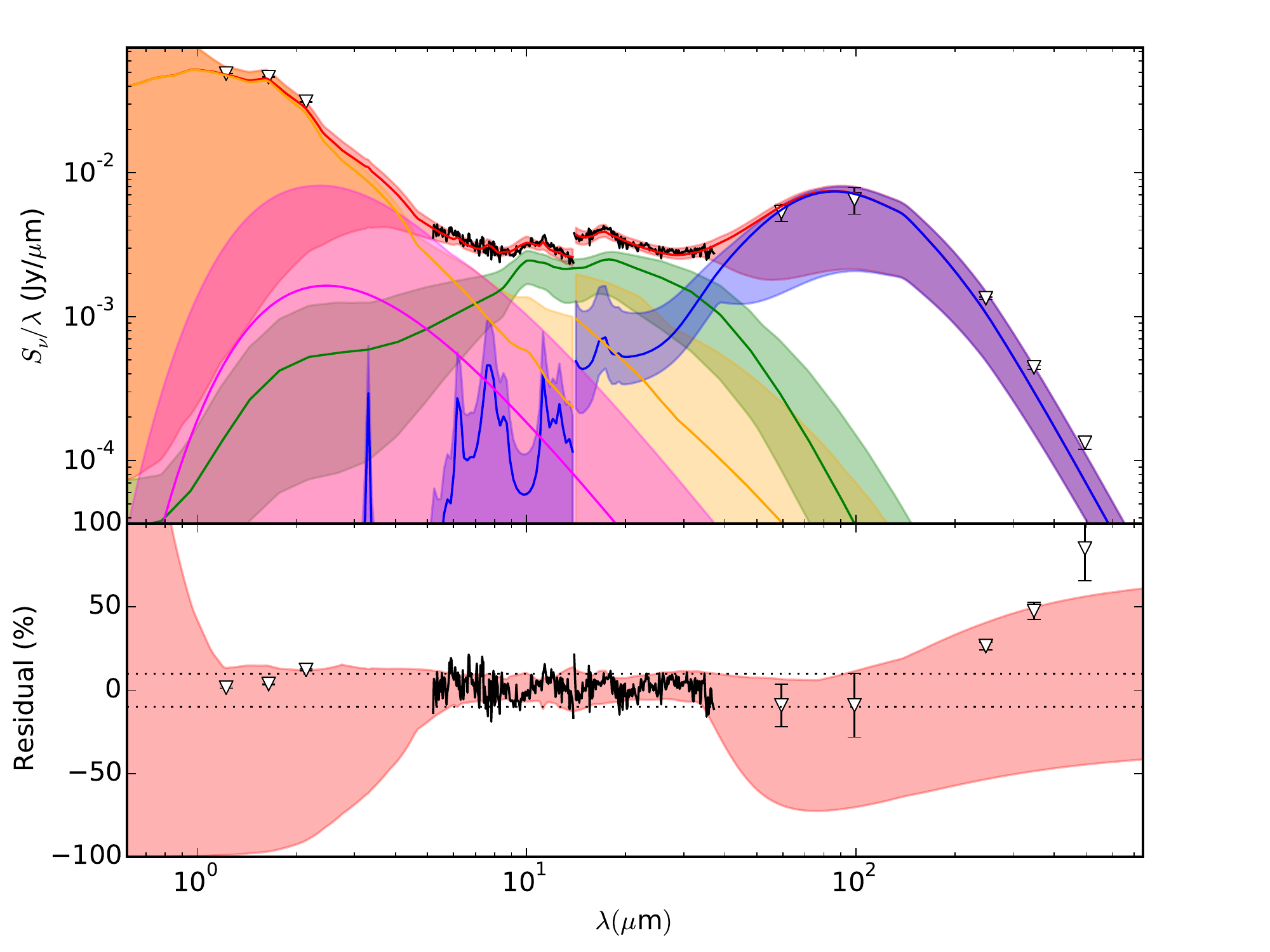}
\caption{SED model-fitting of NGC\,4235. The bestfit model is with a (i) clumpy torus, (ii) diffuse ISM, (iii) stellar population and (iv) a hot dust component. 
Blue (line, band) = (best-fit, acceptable-fit) Diffuse ISM component (Draine \& Li models);
Green = torus component (Nenkova et al. models);
Orange = simple stellar population (large stellar population with short birth-time scale, allowed to age passively) (GRASIL models);
Magenta = hot dust component;
Red = total;
Solid black line = IRS spectra;
Black dots with triangles = 2MASS + IRAC measurements (as available);
Open triangles = upper (pointed down) or lower (pointed up) limits, depending on aperture size relative to IRS.
The residuals plot shows the range of acceptable models after the best-fit model has been subtracted. The dotted lines = $\pm10$\% residuals, roughly the expected systematic uncertainty based on the systematic uncertainty of the models and calibration.}
\label{fig6}
\end{figure}

Observations of the Seyfert galaxies NGC\,4235 and NGC\,4594, at the low radio frequencies of 325 and 610 MHz, tentatively reveal signatures of episodic AGN activity in them.
Both the 610~MHz total intensity and the $325-610$~MHz spectral index images suggest the presence of a ``relic'' radio lobe in the Seyfert galaxy NGC\,4235. Based on a {\it simple} spectral ageing analysis, the relic outer lobe appears to be at least two times older than the present lobe. This implies that the AGN in NGC\,4235 was switched ``off'' for the same time that it has been ``on'' for the current episode. If we assume that the lobe expansion speed was constant for the two activity episodes, the outer lobe should have been at least two times in overall extent than the inner lobe. However, since this is not the case (extent is about 1.5, see Section~\ref{morph} and Figure~\ref{fig1}), we can conclude that the lobe expansion was faster in the inner lobe compared to the outer one. This is plausible if the previous AGN activity episode cleared out the surrounding medium, so that the new lobe propagates in a rarer medium. The detection of the ``relic'' lobe makes NGC\,4235 only the second known Seyfert galaxy after Mrk\,6 to exhibit lobe emission from multiple AGN activity episodes.

A $\sim3$~kpc linear ``spur-like'' feature is observed nearly perpendicular to the double-lobed structure in NGC\,4594, at 325 and 610~MHz. This feature {was} not detected in the 5~GHz VLA image of \citet{Gallimore06}, but appears to have been detected in the 1.5~GHz VLA image of \citet{Bajaja88}, although the authors do not comment on it. Both these observations indicate a spectral index of around $-1.0$ between 610 MHz and the GHz frequencies. It is interesting to note that the Milky Way Spur, which is $\approx$4.0~kpc long and extends perpendicular to the Galactic plane, has been suggested to be a one-sided jet or a magnetic tornado produced by a twisted poloidal magnetic field between the disk and halo \citep{Sofue89}. Could the ``spur'' in NGC\,4594 have a similar origin~? Since the VLBI jet in NGC\,4594 is perpendicular to this linear feature, the AGN does not seem to be currently fuelling it. {Since the new data of \citet{Wiegert15} show a much larger northern ``plume'' along the galaxy minor axis at a PA that is intermediate between the ``spur'' PA and the lobe PA, we wonder if the ``spur'' is part of a larger relic lobe emission in NGC\,4594. However, as of now,} the origin of this peculiar feature remains unclear. 

Constraints on the star formation rates derived through infrared observations, support the suggestion that the radio lobes are AGN- rather than stellar-wind-powered, in both the Seyfert galaxies. The best supporting evidence for the AGN-connection comes from the detection of parsec-scale radio jets in both Seyfert galaxies, with position angles similar to the kpc-scale lobe emission, indicative of a close connection. However, jet curvature, going from parsec to kiloparsec scales, is also implied. Curved jets could be indicative of precessing accretion disks, or instabilities in the radio outflows \citep[e.g.,][]{Pringle97,Hardee82}. \citet{Kharb06} had modelled the entire radio lobe in Mrk\,6 to be the projection of a precessing jet. Edge-brightened lobe-like structures {(or figure-of-eight-shaped structures)} can indeed be produced in hydrodynamical jet simulations of slowly precessing jets (M. Nawaz and G. Bicknell, private communication). Based on the torus inclination derived from the SED modelling of NGC\,4594, the radio jet is not perpendicular to the torus: rather it lies between $\approx8\degr$ and $16\degr$ to the torus axis. This could suggest a warp in the accretion disk, which could in turn explain the S-shaped parsec-scale jet as a precessing jet emerging from the accretion disk in NGC\,4594.

AGN bolometric luminosities have been estimated using the [O {\sc iii}] line fluxes, as well as the IR SED modelling. These are tabulated in Table~\ref{tabbol}. The L$_\mathrm{bol}$ values derived from the [O {\sc iii}] line fluxes are eight (for NGC\,4235) and four (for NGC\,4594) times larger than the L$_\mathrm{bol}$ values derived from the {infrared} SED. This is likely to be due to the large uncertainties associated with the bolometric calibration from [O {\sc iii}] as mentioned in \citet{Heckman04}: the derived bolometric luminosities may be off by as large a factor as 14 (using $3\sigma$ quoted in Heckman et al.). We have therefore used the L$_\mathrm{bol}$ values from SED modelling for all calculations. We obtain Eddington ratios of 0.001 for NGC\,4235 and $\sim1\times10^{-5}$ for NGC\,4594. While the Eddington ratios of NGC\,4235 and NGC\,4594 are typical of those derived in Seyfert galaxies \citep[e.g.,][]{Ho08}, it  is interesting to note that between the two sources, the source with the higher radio-loudness parameter, {\it viz.,} NGC\,4594, has the smaller Eddington ratio. Such a trend has indeed been observed for low luminosity AGN \citep[e.g.,][]{Panessa07,Merloni07}. \citet{Rafter09} have identified a large number of broad-line AGN having radio-loudness parameters between 10 and 100, and referred to them as ``radio-intermediate''. NGC\,4235 and NGC\,4594 fall at the extreme left edge (or beyond) of their log~($R$) versus log~($\lambda_{Edd}$) plot ({their} figure~7), just above the log~$R$ = 1 ``radio-loudness'' dividing line. They are both clearly ``radio-intermediate'' by the definition of \citet{Rafter09} and extend the log~($R$) $-$ log~($\lambda_{Edd}$) analysis to much smaller Eddington ratios.

The kinetic to bolometric power ($Q$/L$_\mathrm{bol}$) is representative of the jet to accretion efficiency: it is $\sim$0.009 for NGC\,4235 and $\sim$0.002 for NGC\,4594. We list the {X-ray} bolometric correction factors ($L_\mathrm{bol}/L_\mathrm{(2-10~keV)}$), the Eddington-scaled bolometric powers (L$_\mathrm{2-10~keV}$/L$_\mathrm{Edd}$), and jet powers ($Q$/L$_\mathrm{Edd}$) {for the two sources} in Table~\ref{tabbol}. The absorption-corrected $2-10$~keV X-ray luminosity $L_\mathrm{(2-10~keV)}$, was obtained from \citet{Liu14}: it is 4.0\,$\times\,10^{41}$~ergs~s$^{-1}$ in NGC\,4235 and 3.8\,$\times\,10^{40}$~ergs~s$^{-1}$ in NGC\,4594. {The X-ray bolometric correction factors tell us how much the X-ray intensity has to be scaled up to obtain the total accretion energy density \citep{Vasudevan07}. These range from $\sim$10 to $\sim$40 for the two galaxies.}

We find that the two Seyfert galaxies follow the same upward linear trend that is observed in the (L$_\mathrm{2-10~keV}$/L$_\mathrm{Edd}$) versus ($Q$/L$_\mathrm{Edd}$) plot of \citet[][figure~2]{Merloni07}, but are not coincident with their best-fit line, even after the X-ray luminosity is multiplied by the bolometric correction of 5 chosen by them. Since all the relevant estimates in this analysis are obtained from the literature, it is difficult for us to identify the reason for, or the significance if any, of this offset. Overall, for NGC\,4235 and NGC\,4594, the bolometric power exceeds the jet power, and is linearly related to it. {This implies that the optical AGN is capable of driving the radio outflow.}

\citet{Ho08} found out that the BLR appears to vanish at the lowest Eddington ratios. This seems to be true in the case of NGC\,4594, which has a very small Eddington ratio of $5\times10^{-5}$ for a black hole mass of $7.9\times10^8$\,M$_{\sun}$ \citep{Dong06}. The IR SED modelling indicates that the extinction due to a dusty torus is not significant in NGC\,4594, as its inclination is nearly face-on. Additionally, the escape probability of photons from the accretion disk/BLR is essentially 100\%. Our data are therefore consistent with NGC\,4594 being a ``true'' type 2 AGN with no torus extinction and no BLR. 

\section{Summary and Conclusions}
Low radio frequency observations with the GMRT reveal unprecedented details in the lobes and galactic disk of two nearby ``radio-loud'' Seyfert galaxies, NGC\,4235 and NGC\,4594. 

\begin{enumerate}
\item
The 610~MHz total intensity image and the $325-610$~MHz spectral index image reveal tentative evidence of a ``relic'' radio lobe in NGC\,4235, pointing to episodic AGN activity in this galaxy. This relic outer lobe is at least two times older than the present lobe. A $\sim3$~kpc linear, steep-spectrum ``spur-like'' feature of unknown origin, is observed nearly perpendicular to the double-lobed structure in NGC\,4594. Galactic disk emission is clearly detected at both radio frequencies in NGC\,4594, but not in NGC\,4235. Galactic contamination makes the flux density and spectral index values for the lobe in NGC\,4594 uncertain.

\item
The infrared SED modelling suggests that the nuclear star formation rate is 0.23~M$_{\sun}$~yr$^{-1}$ and 0.13~M$_{\sun}$~yr$^{-1}$ in NGC\,4235 and NGC\,4594, while the SFR required to power the radio lobes entirely is 2.7~M$_{\sun}$~yr$^{-1}$ and 2.0~M$_{\sun}$~yr$^{-1}$, respectively. This supports the idea that the lobes are AGN, rather than stellar-wind powered. The global SFR estimates from both the infrared and radio data are consistent within the errors for the galactic disk emission in NGC\,4594. 

\item
The lobe-AGN connection is further strengthened by the presence of parsec-scale jets in NGC\,4594 and a sub-kiloparsec-scale radio jet in NGC\,4235, that are roughly at the same position angle as the radio lobes, and are therefore likely to be powering them. The SED model-fitting and VLBI imaging suggests that the parsec-scale jet in NGC\,4594 is inclined at $\approx8\degr$ to $16\degr$ to the torus axis: its S-shape could therefore indicate a precessing jet emerging from a warped accretion disk. A curved jet is also implied in NGC\,4235.

\item
The SED modelling strongly supports the suggestion that NGC\,4594 is a ``true'' type 2 AGN with insignificant dust obscuration and a weak or absent BLR. The dusty torus inclination is 4$\degr$ in NGC\,4235 and 17$\degr$ in NGC\,4594, while the ``escape probability'' for the accretion disk and BLR photons to escape the torus, is essentially $100\%$ in both NGC\,4235 and NGC\,4594, making them both type 1 ({in the sense of being} unobscured) in the Unified scheme. The SED fitting reveals that NGC\,4235 has an additional hot dust component (temperature $\sim1500$~K), apart from the torus, diffuse ISM and stellar components.

\item
AGN bolometric luminosities derived from the [O {\sc iii}] line fluxes are four to eight times larger than the more accurate luminosities derived from the IR SED modelling. The Eddington ratios obtained from the latter are $\sim$0.001 for NGC\,4235 and $\sim1\times10^{-5}$ for NGC\,4594.

\item
The time-averaged kinetic power of the outflow is $3.2\times10^{40}$~ergs~s$^{-1}$ in NGC\,4235. Since galactic emission contaminates the total lobe emission in NGC\,4594, an upper limit of the kinetic power is $2.1\times10^{39}$~ergs~s$^{-1}$. Such kinetic powers have been found in other low-luminosity AGN with direct estimates of mass outflow rates. The relations for kinetic power that are derived for powerful radio galaxies are also valid for Seyfert galaxies with radio lobes. {The jet to accretion efficiency factor, represented by $Q$/L$_\mathrm{bol}$, is $\sim$0.009 for NGC\,4235 and $\sim$0.002 for NGC\,4594. The X-ray bolometric correction factors range from $\sim$10 to $\sim$40 for the two galaxies.}

\item
Between the two Seyfert galaxies, there is an inverse relation between the radio-loudness and Eddington ratio and a direct relation between the Eddington-scaled jet power and bolometric power. Such trends have indeed been found for low luminosity AGN in the literature.
\end{enumerate}

\section*{Acknowledgements}
{We thank the referee for their helpful suggestions which have significantly improved this manuscript.}
We thank the staff of the GMRT that made these observations possible. GMRT is run by the National Centre for Radio Astrophysics of the Tata Institute of Fundamental Research. {This publication makes use of data products from the Two Micron All Sky Survey, which is a joint project of the University of Massachusetts and the Infrared Processing and Analysis Center/California Institute of Technology, funded by the National Aeronautics and Space Administration and the National Science Foundation. This work is based in part on observations made with the Spitzer Space Telescope, which is operated by the Jet Propulsion Laboratory, California Institute of Technology under a contract with NASA. Support for this work was provided by NASA through an award issued by JPL/Caltech.} This research has made use of the NASA/IPAC Extragalactic Database (NED) which is operated by the Jet Propulsion Laboratory, California Institute of Technology, under contract with the National Aeronautics and Space Administration. AH acknowledges Faculty Recharge Programme of the University Grant Commission (India) for the award of Assistant Professor position. 

\bibliographystyle{mnras}
\bibliography{ms}

\onecolumn

\begin{table}
\centering
\caption{Equipartition Estimates for the Lobes}
\begin{tabular}{lccccccccccc}
\hline\hline
{Comp} & {$S_{tot}$} & {Major} & {Minor} & {$\phi$} & {$L_{rad}$} & {$B_{min}$}  & {$E_{min}$} & {$P_{min}$}& {$E_{tot}$} & {$U_{tot}$} & {$t_e$ }\\ 
{} & {} & {} & {}   & {} &{$\times10^{39}$} & {$\times10^{-6}$} & {$\times10^{54}$} & {$\times10^{-12}$} & {$\times10^{54}$} & {$\times10^{-11}$} & {$\times10^6$}\\
{} & {mJy} & {arcsec} & {arcsec} & {} & {ergs\,s$^{-1}$} & {G} & {ergs} & {dynes\,cm$^{-2}$} & {ergs} & {ergs~cm$^{-3}$} & {yr}\\ 
\hline
NGC\,4235$w$ &  8.6 & 52.1 & 47.2  & 1              & 15.8 & 4.5  &  65.8 &  1.9 &  82.3  & 0.4 &     37.3 \\
NGC\,4235$e$  &  4.2 & 43.1 & 40.1  & 1              & 7.8   & 4.3  &  35.2 & 1.7 &   44.0  & 0.4  &    38.4 \\
NGC\,4594$s$  &  5.6 & 36.2 & 20.8  & 1              & 1.9   & 7.9  &  3.4  &  5.8  &  4.2    & 1.2  &    23.1 \\ \hline
NGC\,4235$w$ &  ''     & ''       & ''       & $10^{-3}$ & "      & 32.5 &  3.4  & 98.1 &  4.3   &  21.0 &   3.4 \\
NGC\,4235$e$  &  ''     & ''       & ''       & $10^{-3}$ & "      & 30.8 & 1.8   & 88.0 &  2.3   & 18.9  &   3.7   \\
NGC\,4594$s$  &   ''    & ''       &  ''      & $10^{-3}$ & "      & 57.0 & 0.2   & 301.4 &  0.2  & 64.6 &   1.5 \\ \hline
\end{tabular}\\
{Images at 610 MHz were used for all estimates. A lobe spectral index of $-0.6$ was adopted throughout. Column\,1: Source name followed by lobe (w=west, e=east, s=south). 
Column\,2: Integrated flux density. Columns\,3 \& 4: Major and Minor axes in arcseconds, respectively.
Column\,5: Volume filling factor.
Column\,6: Total radio luminosity. Column\,7: Minimum magnetic field strength. Column\,8: Minimum energy. Column\,9: Minimum pressure. Column\,10: Total energy in particles and fields $E_{tot}$ (= 1.25~$\times~  E_{min}$). Column\,11: Total energy density $U_{tot}$ = E$_{tot}~(\phi~V)^{-1}$. Column\,12: Electron lifetimes in years.}
\label{tabequip}
\end{table}

\begin{table}
\centering
\caption{Eddington-scaled Quantities and Star-formation rates}
\begin{tabular}{lcccccccccc}
\hline\hline
{Source}&{$M_{bh}$}&{$L_{bol}$-1}&{$L_{bol}$-2}&{$L_{Edd}$}&{$\lambda_{Edd}$} &{$L_{bol}$/$L_\mathrm{2-10~keV}$}&{$L_\mathrm{2-10~keV}$/$L_{Edd}$}&{$Q/L_{Edd}$}&{SFR-1} &{SFR-2}\\ 
{} & {$M_{\sun}$} & {ergs\,s$^{-1}$} & {ergs\,s$^{-1}$} & {ergs\,s$^{-1}$} & {} & {} & {} & {} & {$M_{\sun}$\,yr$^{-1}$} & {$M_{\sun}$\,yr$^{-1}$} \\\hline
NGC\,4235&$2.5\times10^7$&$2.9\times10^{43}$&$3.5\times10^{42}$&$3.1\times10^{45}$&0.001                  &9  &0.0001&$1.0\times10^{-5}$&2.7&0.23 \\
NGC\,4594&$7.9\times10^8$&$5.2\times10^{42}$&$1.4\times10^{42}$&$9.9\times10^{46}$&$1.4\times10^{-5}$&37&$3.8\times10^{-7}$&$2.1\times10^{-8}$&2.0&0.13 \\ \hline
\end{tabular}\\
{Column\,1: Source name. Column\,2: Black hole mass from \citet{Dong06}. Column\,3: Bolometric luminosity estimated from [O {\sc iii}] line luminosity. Column\,4: Bolometric luminosity estimated from IR SED modelling. Column\,5: Eddington luminosity. Column\,6: Eddington ratio ($\lambda_\mathrm{Edd}\equiv$L$_\mathrm{bol}$/L$_\mathrm{Edd}$). Column\,7: X-ray bolometric correction factor. Column\,8: Eddington-scaled bolometric power. Column\,9: Eddington-scaled jet power. Column\,10: Star formation rates derived from the radio lobe emission. Column\,11: Nuclear star formation rates derived from the IR SED modelling.}
\label{tabbol}
\end{table}

\begin{figure}
\centering
\includegraphics[width=18cm]{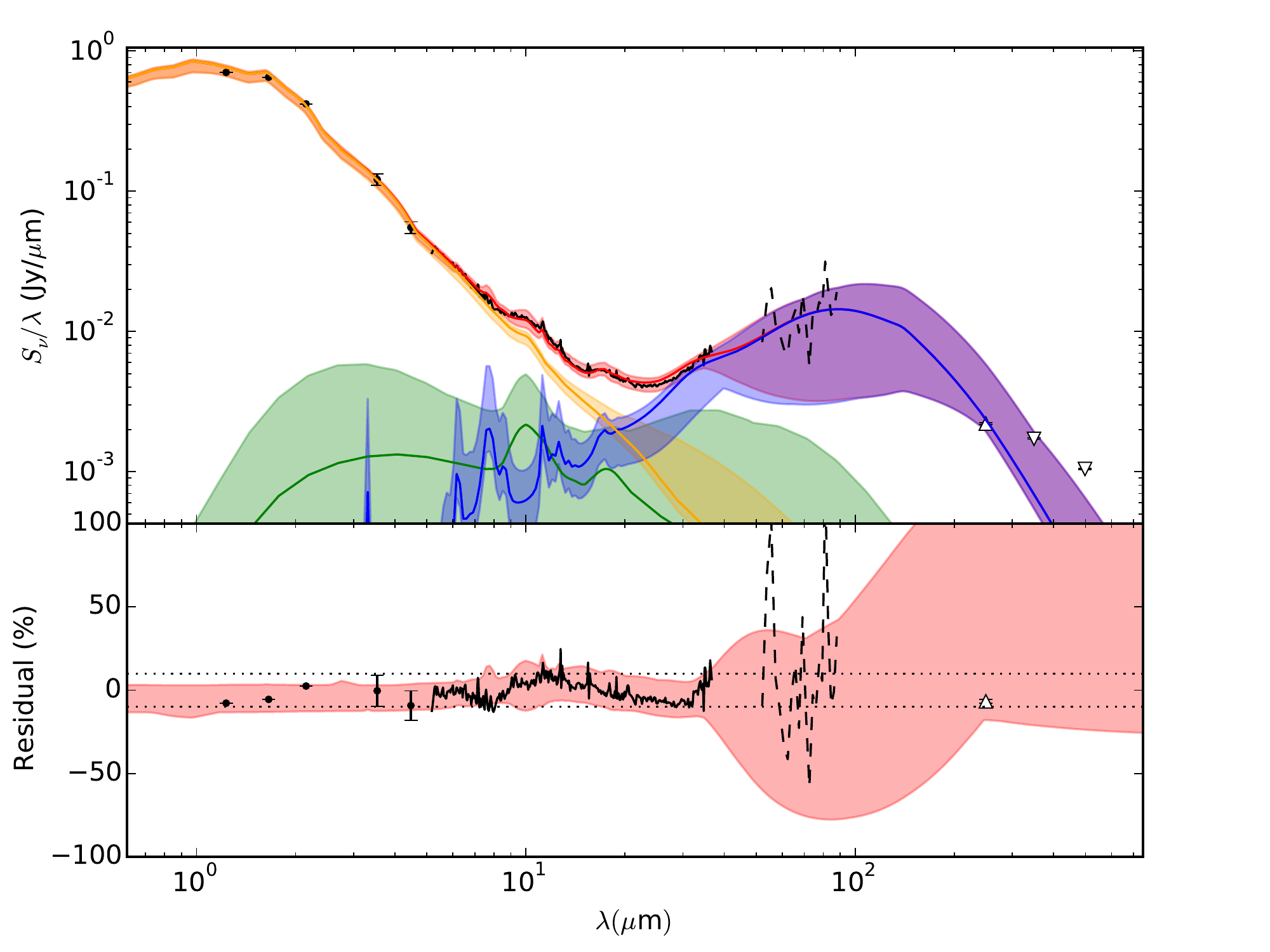}
\caption{SED model-fitting of NGC\,4594. The bestfit model is with a (i) clumpy torus, (ii) diffuse ISM, and (iii) stellar population. For component details, see caption under Figure~\ref{fig6}.}
\label{fig7}
\end{figure}

\begin{table}
\centering
\caption{Best-fit Parameters from SED Modelling for NGC\,4235}
\begin{tabular}{llll}
\hline
{Parameters} & {bestfit} & {median ($-\sigma$, $+\sigma$)} & {($-2\sigma$, $+2\sigma$)}\\ \hline
Torus \citep{Nenkova08} &  &  & \\ \hline
$\sigma$ &  16.0 &  20.0 ($-3.7, 8.4$) & ($-4.8, 22.5$) deg \\
       $Y$ &  20.3 & 19.2 ($-5.4, 16.6$) & ($-9.1, 64.5$) \\
       $N_0$ &  14.7 & 11.8 ($-3.2, 2.2$) & ($-6.5, 3.0$) \\
        $q$  &  1.21  & 1.21 ($-0.72, 0.72$) & ($-1.12, 1.18$) \\
     $\tau_v$ & 138.4 & 205.5 ($-81.5, 65.8$) & ($-139.0, 90.3$) \\
 $i$  &   3.7  & 18.3 ($-13.0, 18.5$) & ($-17.5, 36.4$) deg \\
 $C_\mathrm{frac}$ &  47.2 & 55.2 ($-8.9, 17.1$) & ($-12.1, 33.0$) percent\\
 hoa &  62.1 & 56.6 ($-13.4, 6.0$) & ($-30.8, 8.0$) deg \\
  esc Prob & 100.00 & 99.98 ($-9.20, 0.02$) & ($-67.11, 0.02$) percent \\
  log Mass & $-0.51$ & $-0.49$ ($-0.24, 0.23$) & ($-0.50, 0.44$) M$_{\sun}$ \\
  log Size &  $-0.32$ & $-0.36$ ($-0.17, 0.26$) & ($-0.35, 0.61$) pc \\ \hline
ISM \citep{Draine07} & & &\\ \hline
$u_{min}$ &   5.9 &  5.7 ($-0.3, 0.3$) & ($-0.6, 0.8$) \\
$q_{PAH}$ & 1.71 & 1.51 ($-0.51, 0.73$) & ($-0.90, 1.90$) \\
$f_{PDR}$   &  57.7  & 62.0 ($-3.6, 3.5$) & ($-6.2, 8.6$) percent \\ \hline
Stars \citep[GRASIL;][]{Silva98} & & &\\ \hline
 Age & 12.9  &  8.4 ($-5.0, 4.3$) & ($-7.8, 4.6$) Gyr \\ \hline
Hot dust & & &\\ \hline
     $Cf$     & 33.2 & 54.1 ($-29.3, 33.0$) & ($-46.1, 47.3$) percent \\
     Temp  & 1488.1 & 1398.5 ($-211.5, 142.8$) & ($-435.6, 192.6$) K \\ \hline
Amp & & &\\ \hline
log AGNLum & 42.541 & 42.500 ($-0.145, 0.102$) & ($-0.264, 0.220$) ergs/s \\
log ISMLum & 42.714 & 42.714 ($-0.007, 0.009$) & ($-0.012, 0.018$) ergs/s \\
   log SFR  & $-0.633$ & $-0.633$ ($-0.007, 0.009$) & ($-0.012, 0.018$) M$_{\sun}$/yr \citep{Kennicutt98}\\
SL apcorr & 0.233  & 0.225 ($-0.064, 0.076$) & ($-0.132, 0.163$)\\
\hline
\end{tabular}
\\
{{\bf Torus:} The first six parameters determine the shape of the model torus spectrum.
$\sigma$ = angular scale height of the torus in degrees, $Y$ = ratio of the outer radius to the inner radius of the torus set by sublimation, $N_0$ = average number of clouds along an equatorial sight-line, $q$ = the power law index of the cloud distribution in radius - larger $q$ values imply that clouds are more centrally concentrated, $\tau_v$ = the optical depth at V-band (550 nm) through one cloud in the torus, $i$ = the inclination of the torus relative to the sight-line, with 0 being face-on. The next five parameters are calculated from the torus parameters after the fit is done. $C_\mathrm{frac}$  = the fraction of sky covered in obscuring clouds as viewed from the centre of the torus, hoa = the half-opening angle: this is the polar angle at which the escape probability for accretion disk photons drops by 1/$e$ relative to the poles, twice this angle is the apparent width of an ionisation cone if the torus were viewed edge on, esc Prob = escape probability of photons from the accretion disk or BLR for the best-fit sight-lines - this should be higher for type 1 AGN compared to type 2, log Mass = the mass of atomic + molecular hydrogen of the torus, log Size = an estimate of the outer radius of the torus assuming that the inner radius is set by 1500 K sublimation temperature and the best-fit values for $Y$.
{\bf ISM:} $u_{min}$ = a lower limit to the scaling of the interstellar radiation field (ISRF) relative to the Milky Way value, $q_{pah}$ = mass fraction of PAHs relative to total dust mass, $f_{pdr}$ = fraction of dust luminosity from photo-dissociation regions (i.e. star-forming regions).
{\bf Stars:} Age = age of the model stellar population in Gyr.
{\bf Hot dust:} $Cf$ and Temp are the sky covering fraction and dust temperature of the hot dust component.
{\bf Amp:} Amplitude of the torus component. log AGNLum = logarithm of the AGN bolometric luminosity, based on the scaling of the clumpy torus model needed for a best fit, log ISMLum = logarithm of the ISM luminosity,  log SFR = logarithm of the star formation rate.
SL apcorr: a measure of how much the large scale (LL) model spectrum of the diffuse ISM and stars has to be scaled down to fit the SL data. See Section~\ref{secir} for details.}
\label{tabsed1}
\end{table}

\begin{table}
\centering
\caption{Best-fit Parameters from SED Modelling for NGC\,4594}
\begin{tabular}{llll}
\hline
{Parameters} & {bestfit} & {median ($-\sigma$, $+\sigma$)} & {($-2\sigma$, $+2\sigma$)}\\ \hline
Torus \citep{Nenkova08} &  &  & \\ \hline
$\sigma$               &  23.0 & 25.3 ($-$8.0, 20.2) & ($-10.1$, 40.1) deg \\
$Y$                       &  45.4  & 44.6 ($-$31.7, 37.5) & ($-$38.7, 52.8) \\
$N_0$                   &  1.6   & 2.0 ($-$0.8, 2.8) & ( $-$1.0, 9.1)\\
$q$                       &  1.65  &  2.27 ($-$0.85, 0.52) & ( $-$1.89, 0.70)\\
$\tau_v$               &  36.5  & 65.9 ($-$48.3, 148.0) & ($-$55.1, 220.9)\\
$i$                        &  17.4  & 39.2 ($-$26.9, 30.5) & ($-$37.4, 47.5) deg\\
$C_\mathrm{frac}$& 33.0  &  43.4 ($-$14.5, 19.5) & ($-$22.7, 42.3) percent\\
hoa                        & 69.1  & 62.9 ($-$15.8, 9.4) & ($-$35.8, 14.9) deg\\
esc Prob                & 99.99 & 89.53 ($-$57.91, 10.47) & ($-$85.09, 10.47) percent\\
log Mass                & $-$2.09  & $-$2.27 ($-$0.52, 0.54) & ($-$0.94, 1.26) M$_{\sun}$\\
log Size                  & $-$0.17  & $-$0.22 ($-$0.54, 0.29) & ($-$0.93, 0.48) pc\\ \hline
ISM \citep{Draine07} & & &\\\hline
$u_{min}$             &  3.9  &  3.6 ($-$1.8, 1.2) & ($-$3.2, 2.3)\\
$q_{PAH}$           &  0.51 &  0.63 ($-$0.12, 0.21) & ($-$0.16, 0.51)\\
$f_{PDR}$            & 17.5  & 14.8 ($-$1.9, 2.8) &  ($-$3.7, 15.6) percent\\ \hline
Stars \citep[GRASIL;][]{Silva98} & & &\\ \hline
Age                       & 13.0 &  12.9 ($-$0.2, 0.1) & ($-$0.4, 0.1) Gyr\\ \hline
Amp & & &\\ \hline
log AGNLum        & 42.148   & 42.077 ($-$0.263, 0.256) & ($-$0.489, 0.480) ergs/s\\
log ISMLum         & 42.455    & 42.470 ($-$0.035, 0.041) & ($-$0.175, 0.082) ergs/s\\
log SFR               & $-$0.892 & $-$0.877 ($-$0.035, 0.041) & ($-$0.175, 0.082) M$_{\sun}$/yr \citep{Kennicutt98} \\\hline
\end{tabular}
\\
{See caption under Table~\ref{tabsed1}.}
\label{tabsed2}
\end{table}

\appendix
\section{Probability density functions}
{We include figures showing the MCMC estimate of the marginalized probability distribution functions for each parameter in the SED modelling. These indicate whether a given parameter is well constrained or if it has only upper or lower limits. For example, the stellar population (SP) age for NGC\,4594 is a lower limit but it is essentially unconstrained for NGC\,4235, from inspection of plots \ref{figb} and \ref{figd}. 
The summary statistics for the same are included in Tables~\ref{tabsed1} and \ref{tabsed2}.}

\begin{figure}
\centering
\includegraphics[width=14cm]{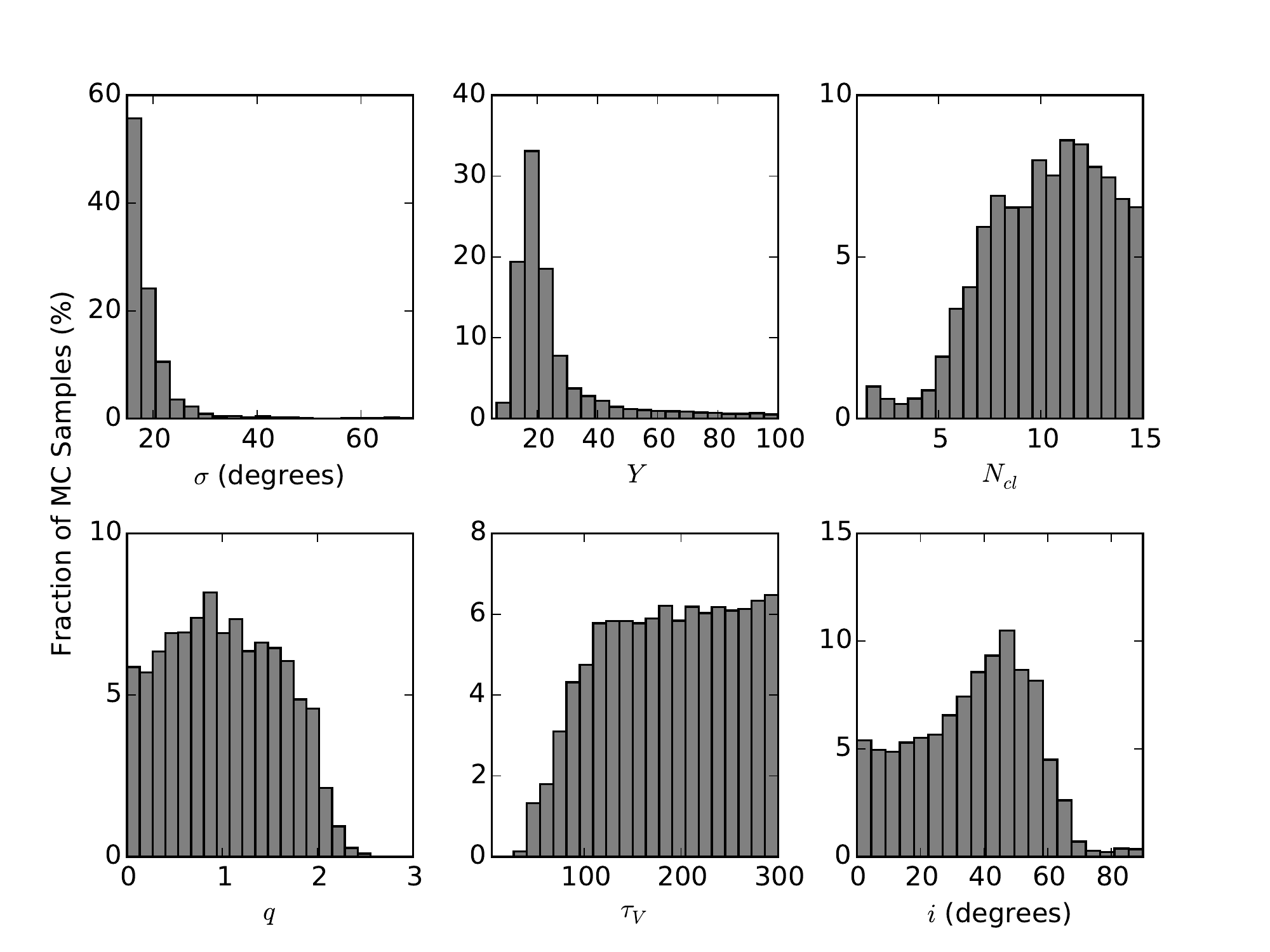}
\includegraphics[width=14cm]{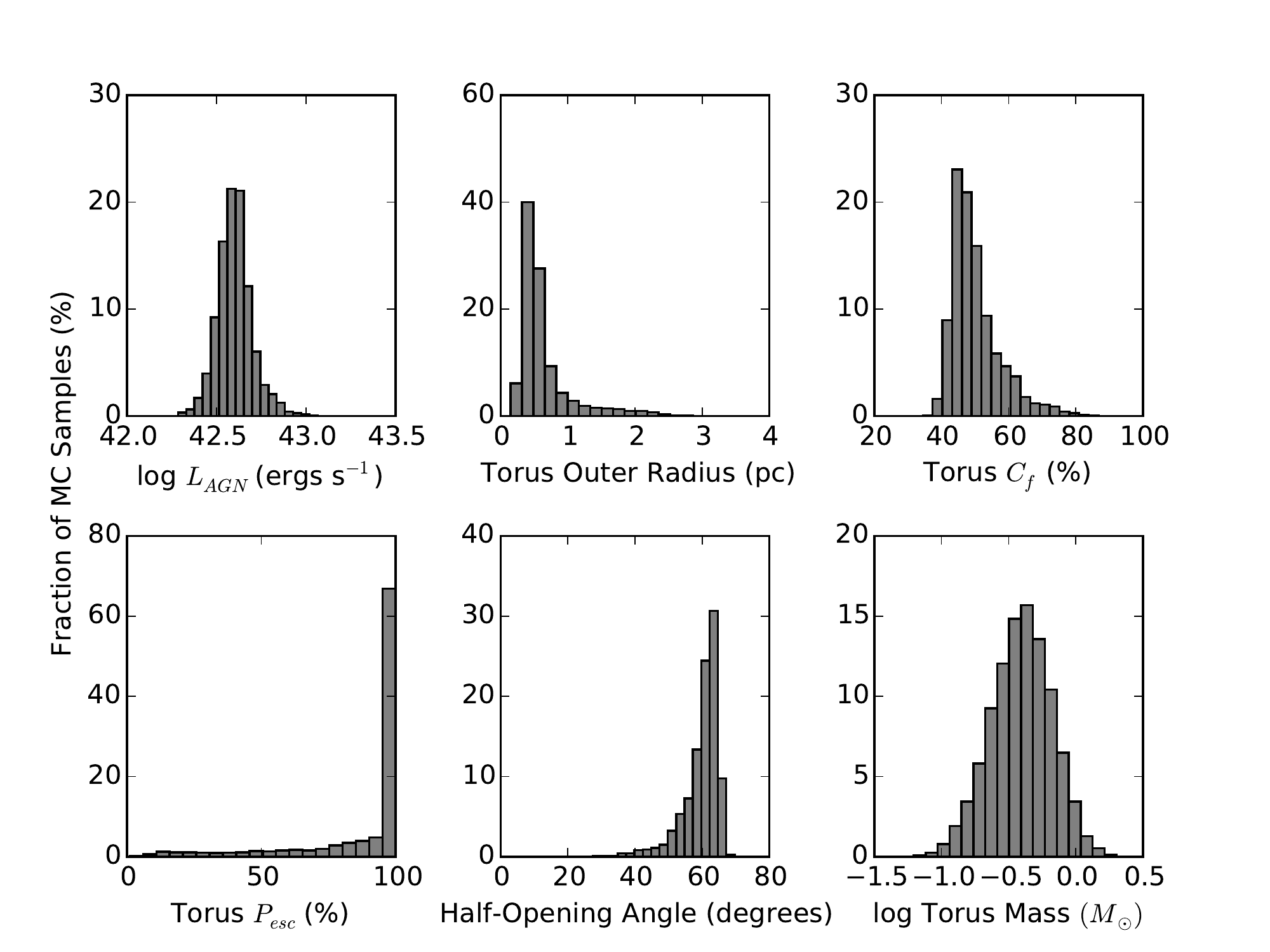}
\caption{Marginalized posterior probability densities estimated from the MCMC samples for the torus parameters derived from the clumpyDREAM SED fits to NGC\,4235.}
\label{figa}
\end{figure}

\begin{figure}
\centering
\includegraphics[width=14cm]{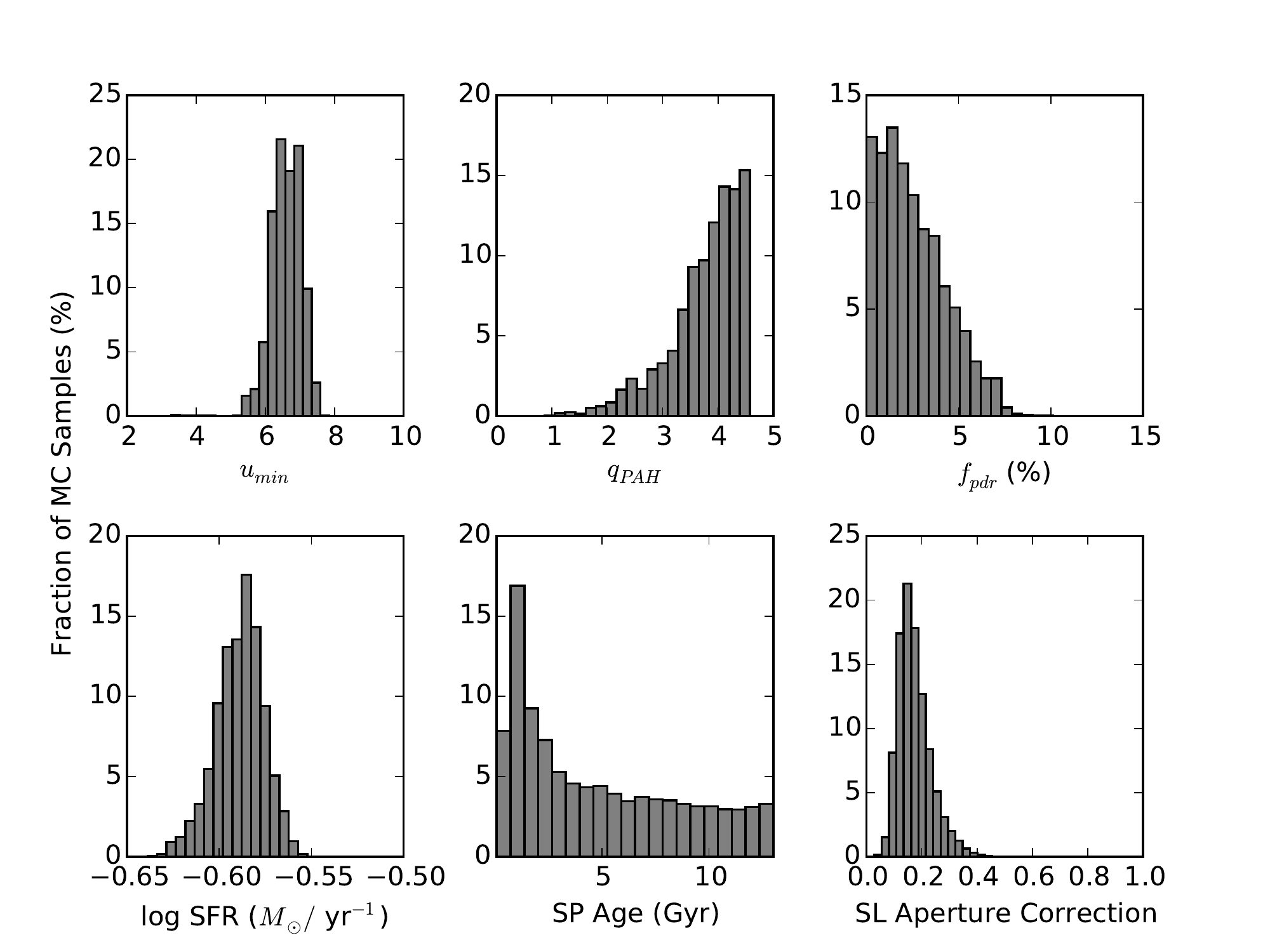}
\caption{Same as Figure~\ref{figa}, for the parameters of the stellar and ISM components.}
\label{figb}
\end{figure}

\begin{figure}
\centering
\includegraphics[width=14cm]{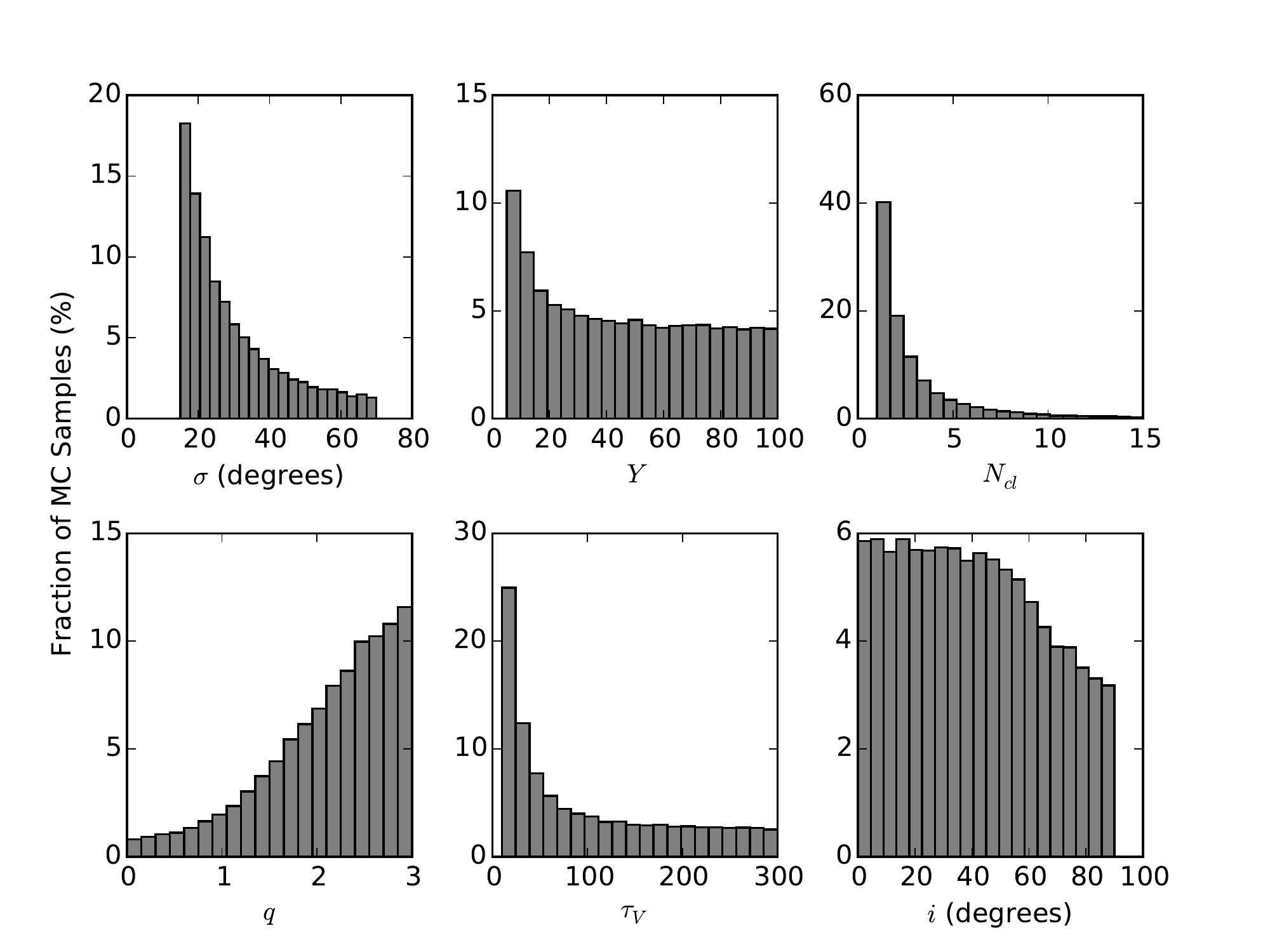}
\includegraphics[width=14cm]{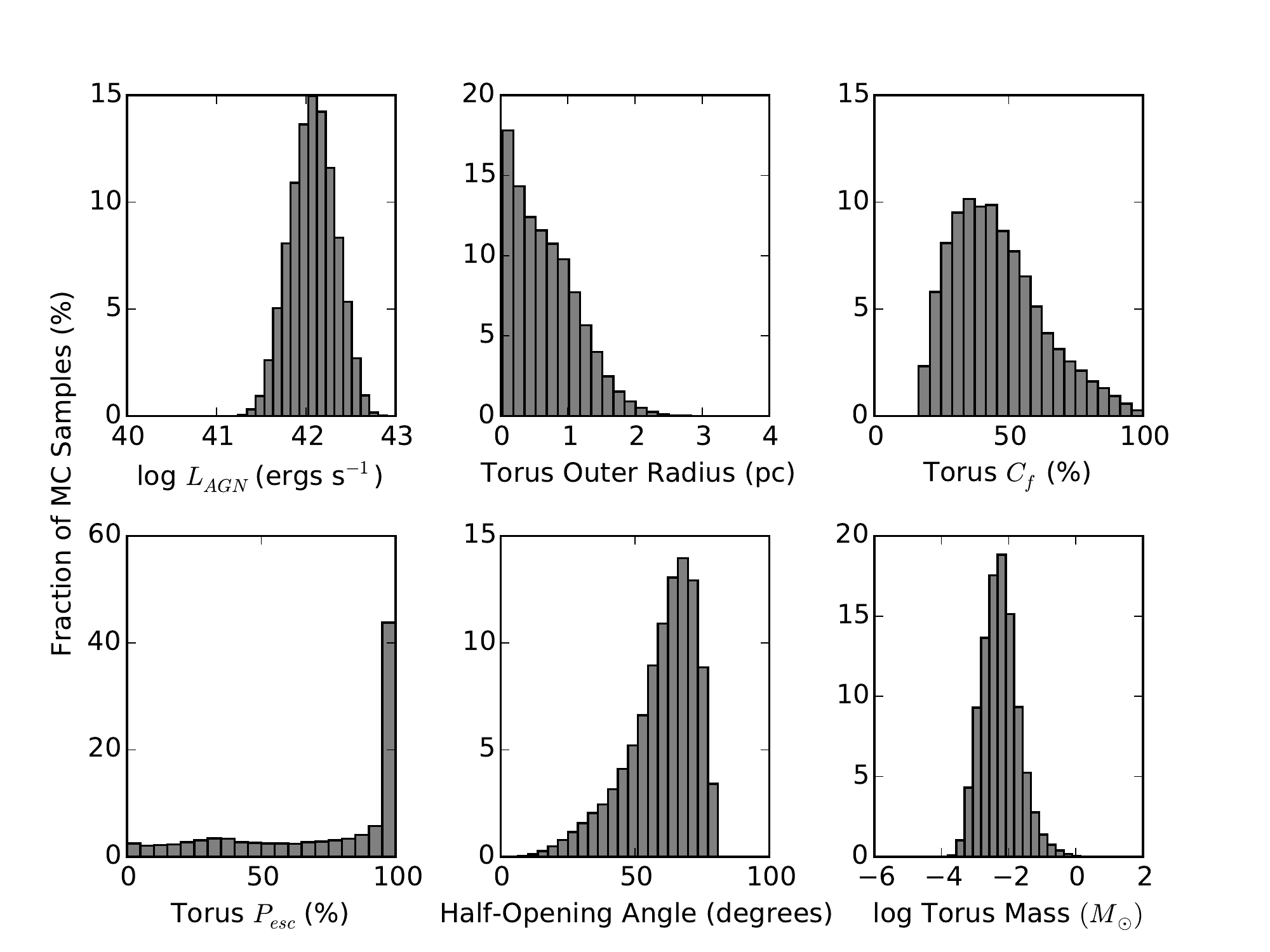}
\caption{Marginalized posterior probability densities estimated from the MCMC samples for the torus  parameters derived from the clumpyDREAM SED fits to NGC\,4594.}
\label{figc}
\end{figure}

\begin{figure}
\centering
\includegraphics[width=14cm]{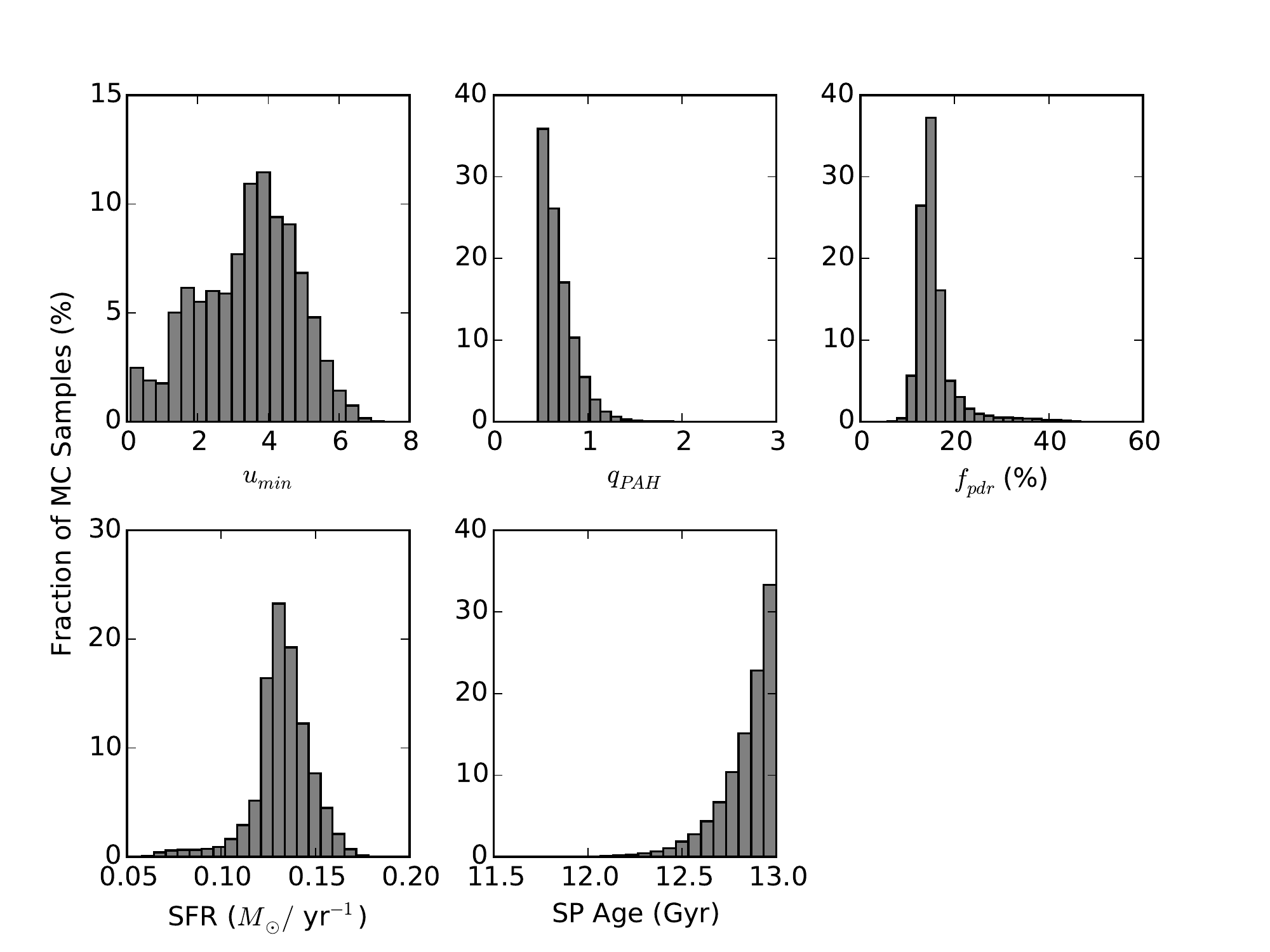}
\caption{Same as Figure~\ref{figc}, for the parameters of the stellar and ISM components.}
\label{figd}
\end{figure}

\bsp	
\label{lastpage}
\end{document}